\documentclass{aa}  

\usepackage{graphicx}   % Including figure files
\usepackage{amsmath}    % Advanced maths commands
\usepackage{caption}
\usepackage{subcaption}
\usepackage{pdflscape}
\usepackage{multirow}
\usepackage{rotating}
\usepackage[dvipsnames]{xcolor}
\usepackage[normalem]{ulem}
\usepackage{booktabs}
\usepackage{placeins}

\usepackage{hyperref}
\makeatletter
\renewcommand*\aa@pageof{,page \thepage{} of \pageref*{LastPage}}
\newcommand{\fink}{{\sc Fink}}

\newcommand{\rev}[1]{\textcolor{black}{#1}}
\newcommand{\revv}[1]{\textcolor{black}{#1}}
\newcommand\footnoteref[1]{\protected@xdef\@thefnmark{\ref{#1}}\@footnotemark}

\newcommand\Tstrut{\rule{0pt}{2.6ex}}         % = `top' strut
\begin{document}

   \title{Transient classifiers for Fink}
\subtitle{Benchmarks for LSST}
   %\subtitle{A close-up into the machine learning classifiers within the Fink broker}
\author{B.M.O. Fraga \inst{\ref{cbpf}}\thanks{\email{bernardo@cbpf.br}}
\and 
C.R. Bom\inst{\ref{cbpf}}$^,$\inst{\ref{cefet}}
\and
A. Santos\inst{\ref{cbpf}}
\and
E. Russeil\inst{\ref{clermont}}
\and 
M. Leoni\inst{\ref{paris}}
\and
J. Peloton\inst{\ref{paris}}
\and
E.E.O. Ishida\inst{\ref{clermont}} 
\and 
A. M\"oller\inst{\ref{swinburne}}$^,$\inst{\ref{ozgrav}}
\and
S. Blondin\inst{\ref{marseille}}$^,$\inst{\ref{eso}}%$^{,}$\inst{7}
}

\institute{Centro Brasileiro de Pesquisas F\'isicas, Rua Dr, Xavier Sigaud 150, Rio de Janeiro, Brazil \label{cbpf}%\email{bernardo@cbpf.br}
\and
Centro Federal de Educa\c{c}\~{a}o Tecnol\'{o}gica Celso Suckow da Fonseca, Rodovia M\'{a}rcio Covas, lote J2, Itagua\'{i},~Brazil \label{cefet}
\and
Universit\'e Clermont-Auvergne, CNRS, LPCA, 63000, Clermont-Ferrand, France\label{clermont}  
\and
Université Paris-Saclay, CNRS/IN2P3, IJCLab, 15 rue Georges Clemenceau, 91405 Orsay, France\label{paris}
\and
Centre for Astrophysics and Supercomputing, Swinburne University of Technology, Mail Number H29, PO Box 218, 31122 Hawthorn, VIC, Australia\label{swinburne}
\and
ARC Centre of Excellence for Gravitational Wave Discovery (OzGrav), John St, Hawthorn, VIC 3122, Australia\label{ozgrav}
\and
 Aix Marseille Univ, CNRS, CNES, LAM, Marseille, France\label{marseille}
 \and
 European Southern Observatory, Karl-Schwarzschild-Straße 2, Garching, D-85748, Germany\label{eso}
 }

   \date{}

  \abstract
   {The upcoming Legacy Survey of Space and Time (LSST) at the Vera C. Rubin Observatory is expected to detect a few million transients per night, which will generate a live alert stream during the entire ten years of the survey. This stream will be distributed via community brokers whose task is to select subsets of the stream and direct them to scientific communities. Given the volume and complexity of the anticipated data, machine learning (ML) algorithms will be paramount for this task. 
   }
   %%%%%
   {We present the infrastructure tests and classification methods developed within the {\sc Fink} broker in preparation for LSST. This work aims to provide detailed information regarding the underlying assumptions and methods behind each classifier and enable users to make informed follow-up decisions from {\sc Fink} photometric classifications.}
  %%%%%%%%%%%%%%
   {Using simulated data from the Extended LSST Astronomical Time-series Classification Challenge (ELAsTiCC), we showcase the performance of binary and multi-class ML classifiers available in \fink. These include tree-based classifiers coupled with tailored feature extraction strategies as well as deep learning algorithms. Moreover, we introduce the CBPF (Centro Brasileiro de Pesquisas F\'isicas) Alert Transient Search (CATS), a deep learning architecture specifically designed for this task. 
   }
  %%%%%%%%%%%%%%%%%%%%%%%
   {Our results show that {\sc Fink} classifiers are able to handle the extra complexity that is expected from LSST data. CATS achieved $\geq 93\%$ \rev{precision for all classes except `long' (for which it achieved $\sim 83\%$)}, while our best performing binary classifier achieves $\geq 98\%$ \rev{precision and $\geq 99\%$ completeness} when classifying the periodic class.}
  %%%%%%%%%%%%%
   {ELAsTiCC was an important milestone in preparing the {\sc Fink} infrastructure to deal with LSST-like data. Our results demonstrate that {\sc Fink} classifiers are well prepared for the arrival of the new stream, but this work also highlights that transitioning from the current infrastructures to \textit{Rubin} will require significant adaptation of the currently available tools. This work was the first step in the right direction.
   }

   \keywords{methods: data analysis --
                surveys --
                supernovae:general
               }

\maketitle
%
%-------------------------------------------------------------------

\section{Introduction}
\label{sec:intro}

The advent of large-scale sky surveys has forced astronomy to enter the era of big data, with current experiments already producing data sets that challenge traditional analysis techniques \citep{hilbe2014}. In this context, machine learning (ML) methods are almost unavoidable \cite[e.g. ][]{bamford2009galaxy,2019arXiv190407248B,bom2022developing}. For time-domain astronomy, the ability to quickly process data and obtain meaningful results has become critical due to current and upcoming projects such as the Zwicky Transient Facility \citep[ZTF;][]{ztf} and the Vera C. Rubin Observatory Legacy Survey of Space and Time \citep[LSST; ][]{lsst_overview}, respectively. As will be the case for LSST, the ZTF project employs a difference imaging analysis pipeline that streams to community brokers, in the form of alerts, every detection above a given signal-to-noise threshold. 

Brokers are subsequently tasked with filtering and analysing the data in detail, selecting the most promising objects for different science cases, and redirecting them to different research communities. {\sc Fink} \citep{fink} is one of the official LSST brokers that has been selected to receive the raw alert stream from the beginning of LSST's operations,\footnote{The other selected brokers are \texttt{ALERCE} \citep{alerce}, \texttt{AMPEL} \citep{ampel}, \texttt{ANTARES} \citep{antares}, \texttt{Babamul}, \texttt{LASAIR} \citep{lasair}, and \texttt{Pitt-Google}.} which are expected to start in 2025. In the meantime, broker systems are operating and being tested with alerts from ZTF. {\sc Fink} ingests and processes the stream, making use of several different science modules that contain cross-matching capabilities, ML classifiers, and user-specified filters \citep[for details on {\sc Fink} see][and references therein]{fink}.   

The experience accumulated in the past few years in {\sc Fink} with ZTF has been paramount for the design, development, and fine-tuning of the broker services according to the needs of different scientific communities. Beyond the scientific results already reported \citep[see, e.g.,][]{grandma, kuhn2023,carry2024,karpov2022}, this partnership has enabled the development of a series of tools specifically designed to deal with the alert stream \citep{leoni2022,biswas2023,russeil_agn,allam2023,lemontagner2023, biswas2023b, karpov2023,Moller:2022ICML}. Nevertheless, given the volume and complexity of the expected data, restructuring algorithms to transition from ZTF to LSST is a non-trivial task. 

\rev{Fortunately, there have been similar situations in the past from which important lessons were learned. The astronomical transient community has a long-standing tradition of preparing for the arrival of data from a new survey by producing detailed light curve simulations and hosting data challenges. The SuperNova Photometric Classification Challenge \citep[SNPCC; ][]{kesslerSNPCC} consisted of simulated data representing three supernova classes (Ia, Ibc, and II) whose light curve properties and rate were built to mimic the then upcoming Dark Energy Survey\footnote{\url{https://www.darkenergysurvey.org/}} \citep[DES;][]{des_review}. The challenge was open to professional astronomers, and received answers from ten different research groups. Despite not identifying a single classifier as being significantly better than the others and even when considering the undeniable differences between the generated simulations and the data finally observed by DES, the challenge was of crucial importance to prepare the necessary tools for the upcoming telescope.}

\rev{This successful experience led to the development of a second simulated data set, this time in preparation for the arrival of LSST. The Photometric LSST Astronomical Time-series Classification Challenge\footnote{\url{https://www.kaggle.com/c/PLAsTiCC-2018}} \citep[PLAsTiCC, ][]{plasticc} was proposed to the general public and received more than 1000 submissions. The data set enclosed 14 different transient classes in the training sample and 15 in the target sample.\footnote{The extra class was intended to test the resilience of algorithms to anomalous observations.} Many algorithms were able to achieve similar results, while the numerically best ranked solution used data augmentation and boosted decision trees \citep{avocado}. Beyond the official results presented within the time frame of both competitions, the most enduring legacy of both challenges has been the data sets they produced, which were subsequently used for the development of numerous methods and tools \citep[e.g. ][ to cite a few]{Alves2022, qu2022,malz2023,allam2023}. }

\rev{All of these efforts constitute the astronomical branch of ML exploring `transfer learning' \citep{transflearning, eriksen2020}, a strategy commonly employed when labels are rare or expensive. It consists of training ML models in one domain (simulations) and using it to leverage information from the target one (real data). The most recent sample of photometrically classified supernova Ia from DES \citep{Moller:2024} was identified following this approach \citep{Moller:2020}.}

\rev{Nevertheless, in all the previous attempts, simulations were generated in the format of full static light curves. The task of on-the-fly classification of nightly detected transient candidates had yet to be approached}. The Extended LSST Astronomical Time-Series Classification Challenge\footnote{\label{foot:elasticc}\url{https://portal.nersc.gov/cfs/lsst/DESC_TD\_PUBLIC/ELASTICC/}} \citep[ELAsTICC;][]{knop2023}, developed by the LSST Dark Energy Science Collaboration (DESC) was created to allow for a concrete assessment of the performance of different stages of the communication pipeline. This includes simulating an alert stream to be received by brokers, the ingestion and analysis of the alerts using ML-based classifications by the broker teams, and reporting back such scores to DESC. \rev{Recently, one of the instances of this data set was used by \citet{elasticc_alerce} to test a multi-class classifier using transformers and by \citet{khakpash2024} in the study of stripped-envelop supernovae.} In this work, we use the first version of the ELAsTiCC data set, streamed between September 2022 and January 2023\footnoteref{foot:elasticc} \rev{in alert format}, to stress test the performance of the entire {\sc Fink} infrastructure in an LSST-like data scenario.

This paper is organised as follows: In Sect. \ref{sec:elasticc}, we present the data set, its properties, and our chosen experiment design. Sect. \ref{sec:exp_design} presents with more details how \fink \space works, showing the preparations for LSST. Sect. \ref{sec:metric} presents the metrics we used to evaluate the classifiers. Sect. \ref{sec:models} gives more details about the construction and training of ML models, while Sect. \ref{sec:class_perform} presents an evaluation of these models on a blind test set. We show how some classifiers can be combined to improve stand-alone performances in Sect. \ref{sec:combined}, and Sects. \ref{sec:discussions} and \ref{sec:conclusions} contain discussions and our conclusions, respectively.

%%%%%%%%%%%%%%%%%%%%%%%%%%%%%%%%%%%%%%%%%%%%%%%%%
%%%%%%%%%%%%%%%%%%%%%%%%%%%%%%%%%%%%%%%%%%%%%%%%%

\section{The ELAsTiCC data set}
\label{sec:elasticc}

The Extended LSST Astronomical Time-series Classification Challenge \citep{knop2023} was designed to test broker systems and classification algorithms when applied to a state-of-the-art data set that mostly resembles LSST alerts. It emerged from the experiences accumulated during two  previous challenges, SNPCC \citep{kesslerSNPCC} and PLAsTiCC\footnote{\url{https://www.kaggle.com/c/PLAsTiCC-2018}} \citep{plasticc}, and it is lead by LSST DESC. Its first objective was to test the  brokers infrastructure capability of ingesting and processing a real-time alert stream. The second goal was to enable the evaluation of ML classification algorithms.

In the first instance of the challenge (hereafter ELAsTiCCv1), two different simulated data sets were provided: a static set of full light curves (one light curve per astrophysical source) was made available in 18 May 2022 and an alert stream corresponding to three years of LSST operations was streamed between September 2022 and January 2023. Both data sets were simulated using SuperNova ANAlysis package \citep[SNANA;][]{kessler2009snana} and contained 19 classes divided into five broad categories\footnote{\label{foot:taxonomy}Full taxonomy can be found at \url{https://github.com/LSSTDESC/elasticc/blob/main/taxonomy/taxonomy.ipynb}} (SN-like, periodic, non-periodic, long, and fast). Light curves, comprising detections and forced photometry in the LSST broad-band filters $\{u,g,r,i,z,Y\}$ were provided (for details on how the simulation was generated we refer to the project  website\footnoteref{foot:elasticc}). In order to isolate the performance of our classifiers, avoid issues due to the data generating process and circumvent the need to transform full light curves into alerts, we chose to use here only the streamed data set.\footnote{A second version, ELAsTiCCv2, was released in mid-2023, with updates in photometric redshift model and cadence. The second version was not used in this work.} \rev{In this format, each alert holds the photometric information obtained at a given day, as well as the previous photometric history of that point in space. Thus, at each new light curve point, a new alert is generated that differs in one point from previous alerts from that same source.}

We selected \rev{one third of all unique objects as training sample for all our algorithms (1\,417\,375 distinct objects corresponding to 17\,214\,758 alerts). The remaining objects were used as a test set to evaluate the performance of our classifiers (2\,874\,008 distinct objects corresponding to 34\,891\,855 alerts).}

\begin{figure*}
    \centering
    \includegraphics[width=\textwidth]{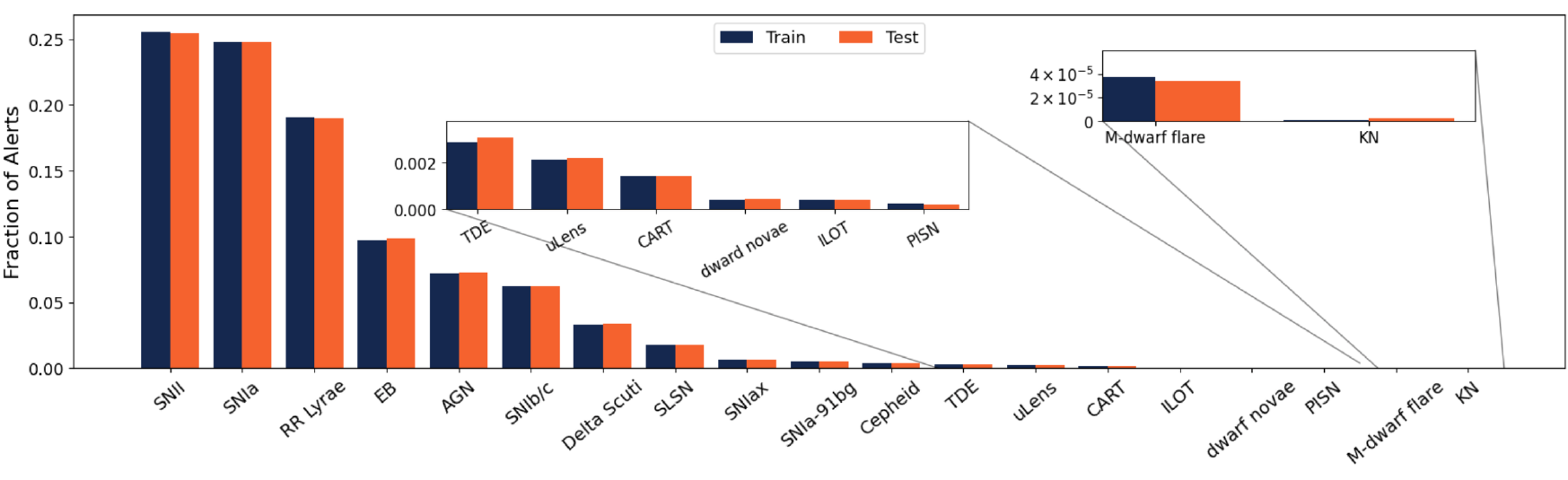}
    \caption{ELAsTiCC class distribution for our training (dark blue) and test (orange) sets.}
    \label{fig:elasticc_train_class_dist}
\end{figure*}

\begin{figure}
    \centering
    \resizebox{\hsize}{!}{\includegraphics{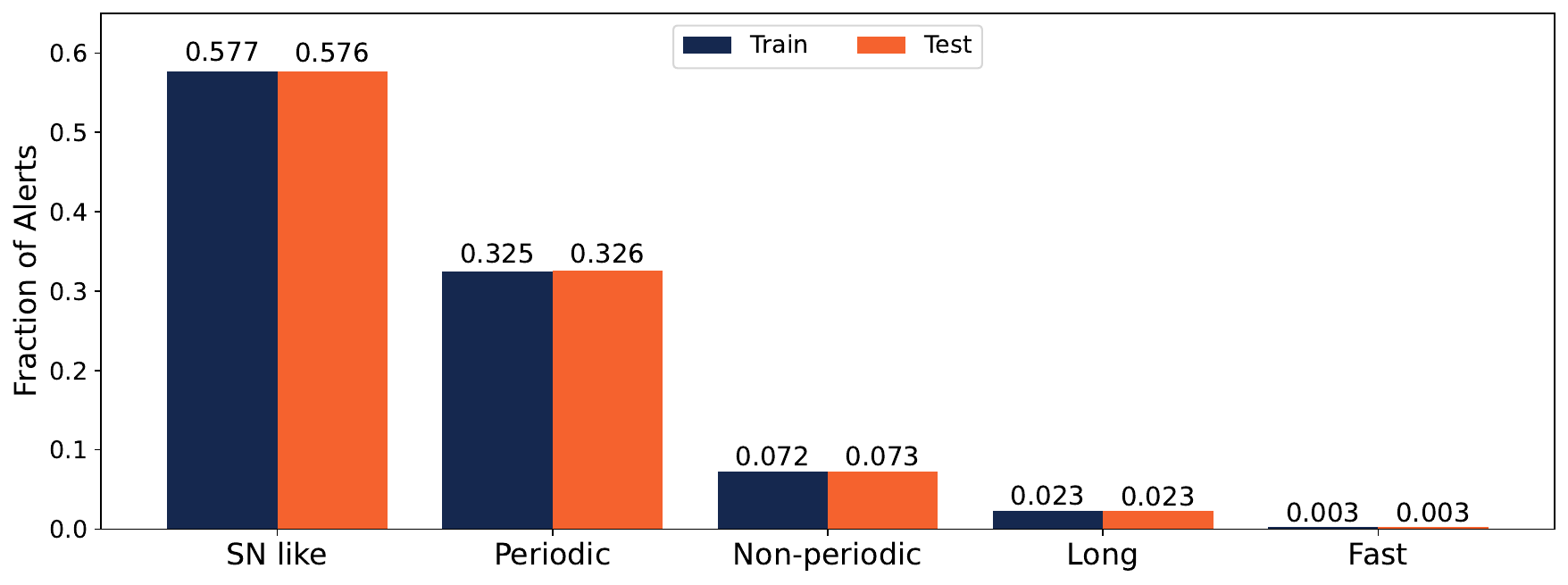}}
    \caption{ELAsTiCC broad class distribution for our training (dark blue) and test (orange) sets.}
    \label{fig:elasticc_train_broadclass_dist}
\end{figure}

The class distributions for the alerts in our train and test sets are shown in Figs. \ref{fig:elasticc_train_class_dist} (unique classes) and  \ref{fig:elasticc_train_broadclass_dist} (broad classes). We note that in our experiment design, the population distributions between classes are similar in both samples, with supernova types Ia (SNIa) and II (SNII) comprising almost half of the alerts, and the fast class being the least represented one. Overall, the distributions for the unique and broad classes are extremely imbalanced, which could pose a problem for multi-class classifiers. 

\begin{figure*}
    \centering
    \includegraphics[width=.49\linewidth]{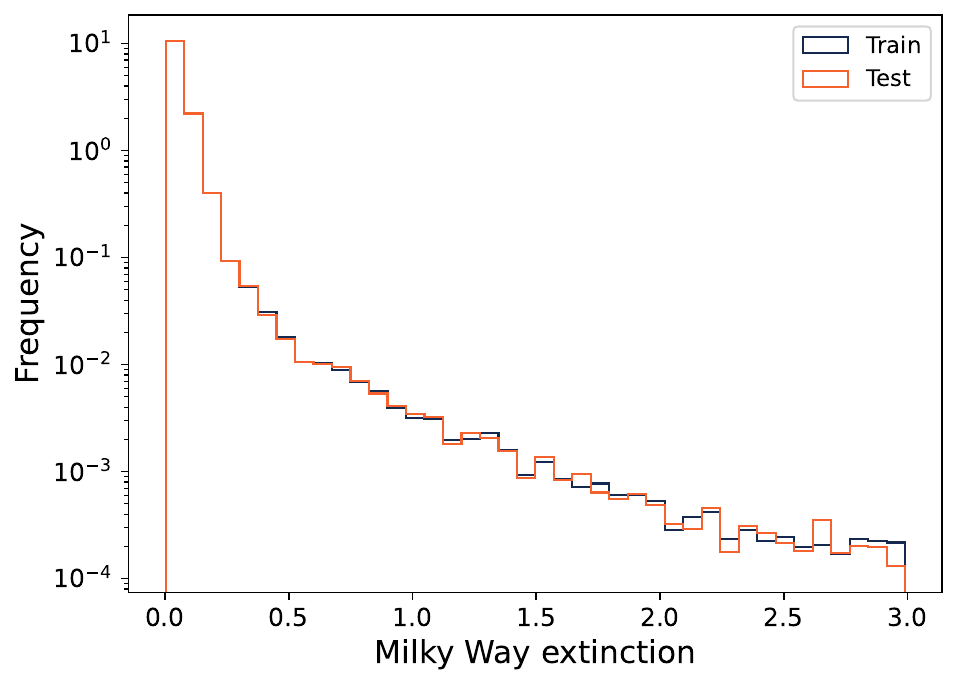} 
    \includegraphics[width=.49\linewidth]{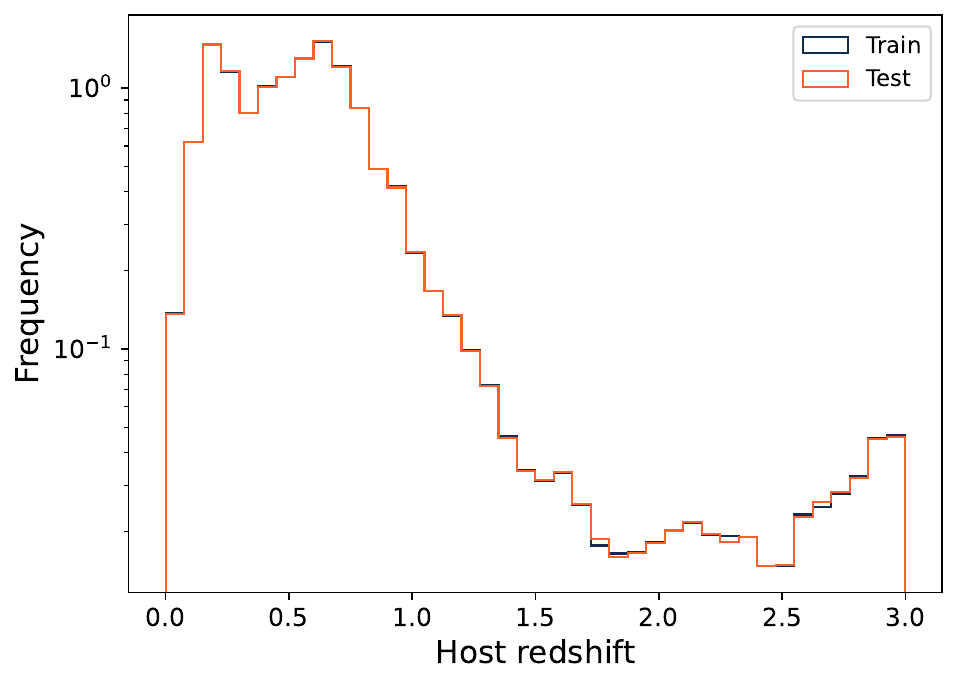}
    \caption{Distribution of milky way extinction (left) and host galaxy photometric redshift (right) for the training (dark blue) and test (orange) sets.}
    \label{fig:meta_dist}
\end{figure*}

Each alert package included both, light curve data (\texttt{mjd}, \texttt{fluxcal}, \texttt{fluxcal\_err}, band-pass filter: \texttt{filtername}) as well as object metadata, comprised of properties such as position, Milky Way extinction and estimated photometric redshift, among others. In Fig. \ref{fig:meta_dist} we show the distributions for two of these %unique object
properties for the training and test set: the Milky Way extinction (\texttt{mwebv}, left) and photometric host galaxy redshift (\texttt{hostgal\_zphot}, right). Approximately \rev{$91\%$ of the objects in the training and test sets have a photometric redshift available.} The distributions are similar, with both photometric redshift distributions displaying a double peaked structure. 

\begin{figure*}
    \centering
    \resizebox{\hsize}{!}{\includegraphics{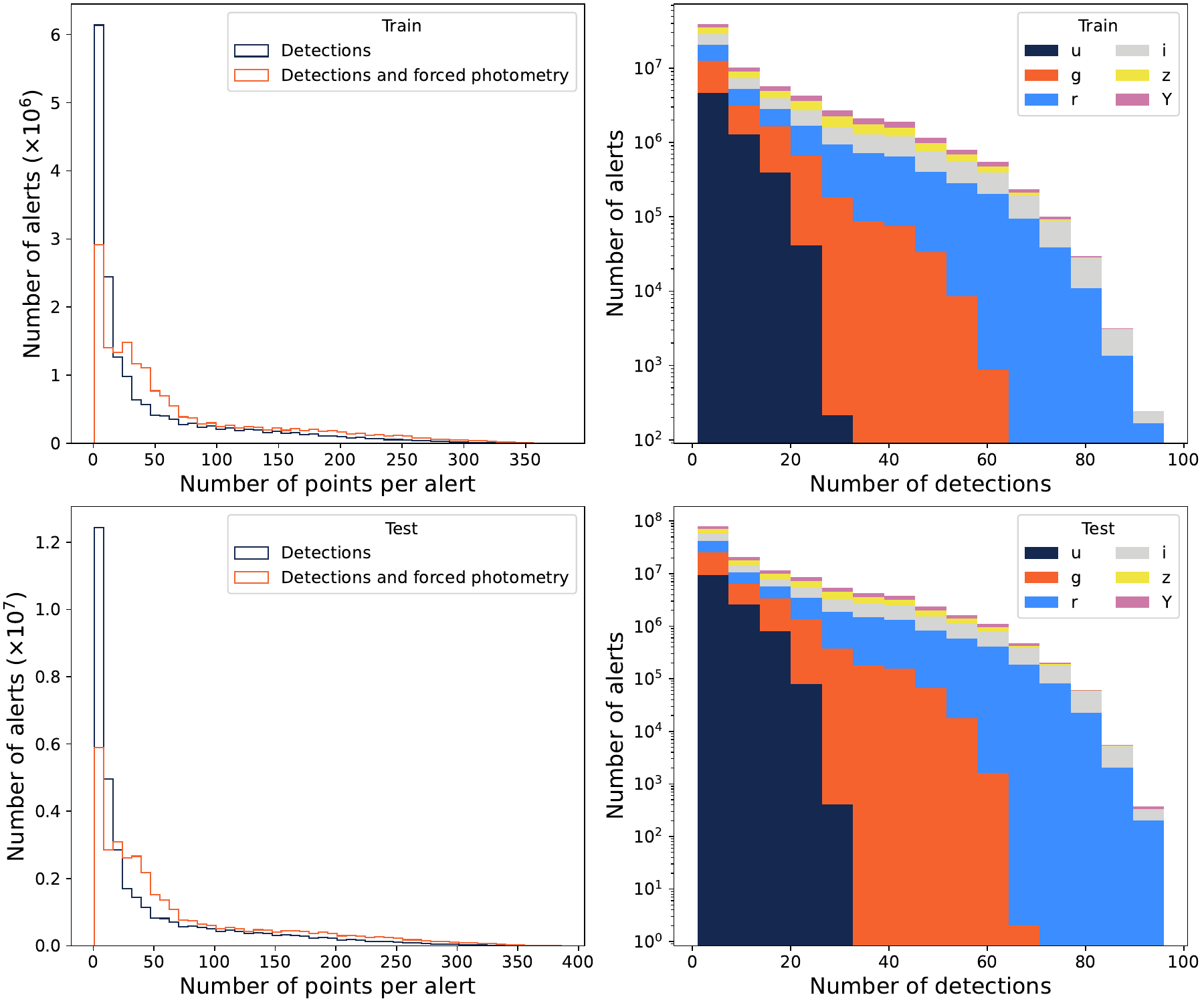}}
    \caption{Distribution of light curve length in the  training (top) and test (bottom) samples. On the left, the total number of points including only detections (blue) and with forced photometry (orange). On the right, the distribution of detections per filter.}
    \label{fig:training_points_perband}
\end{figure*}

Fig~\ref{fig:training_points_perband} shows the distribution of number of detection points per alert with and without forced photometry (left), as well as the global number of detection points in each bandpass (right), for the training (top) and test (bottom) sets. It can be seen from the left panel that the distribution of light curve sizes considering only the detections is strongly peaked around 10 detections, dropping heavily after that, with very few alerts having more than \rev{50 measurements in both the training and test sets. Including forced photometry, the peak is still present around 10 detections, although less pronounced; the decrease is also less steep, as expected. Overall, the distribution is similar for both sets, with the majority of alerts having less than 50 points, even including forced photometry.} 

The distribution of detections per passband (right column of Fig. \ref{fig:training_points_perband}) is similar for the training and test sets: the redder the band, the larger the maximum number of detections. The exception to this is the $z$ band, which do not appear in longer light curves. This is especially important for classifiers that rely on colours or use only specific passbands. Nevertheless, this feature is a direct consequence of the chosen survey strategy, and it is reasonable to expect that the real data will also hold differences in number of detections on each band. Thus, it is paramount to assess the robustness of classifiers in this scenario. 
%%%
%%%%%%%%%%%%%%%%%%%%%%%%%%%%%%%%%%%%%%%%%%%%%%%%%
%%%%%%%%%%%%%%%%%%%%%%%%%%%%%%%%%%%%%%%%%%%%%%%%%
\section{Fink infrastructure}
\label{sec:exp_design}
%%%%%%%%%%%%%%%%%%%%%%%
Since 2019, {\sc Fink} has been processing alert data from the ZTF public stream. Not only the ZTF data rate is lower than the one intended for LSST, but the schema and content of the alert packets from each stream are also different. Therefore, the ELAsTiCC challenge was an opportunity to show that the Fink architecture can scale in terms of data volume and that it is possible adapt the current classifiers, or create new ones, to a new schema with different incoming information.

{\sc Fink} operates in real time on large-scale computing infrastructures. For the ZTF processing, the incoming data stream is provided by the ZTF Alert Distribution System \citep[ZADS;][]{Patterson_2019} that runs a Kafka instance on the DigitalOcean cloud infrastructure (USA). {\sc Fink} is currently deployed on the VirtualData\footnote{\url{https://virtualdata.fr/}} cloud infrastructure (France), and makes use of distributed computing to split the incoming stream of alerts into smaller chunks of data to be analysed independently over many machines in parallel. For LSST, {\sc Fink} will be deployed at CC-IN2P3 (Centre de Calcul de l'Institut National de Physique Nucl\'eaire et de Physique des Particules), which will have a local copy of LSST data that can be efficiently exploited for internal cross-match needs.

What is deployed for ZTF typical rates (of the order of 200,000 alerts per night can be easily scaled to ELAsTiCC rates (of the order of 1,000,000 alerts in a few hours every night) by adding more machines. In practice for the challenge, we use a total of 33 cores for all operations (listening to the incoming stream, processing it and sending back results). Moreover, the processing was done on the same platform, alongside the real-time processing of ZTF alerts.

For the ELAsTiCC challenge, {\sc Fink} operations consisted in three steps,  done in real time: (1) decoding the incoming alert stream, (2) applying all classifiers on every alert, and (3) redistributing all enriched alert packets to the DESC team in the USA. The first and last steps are under the responsibility of the {\sc Fink} engineering team, while the second part involves classifiers provided by the community of users. \rev{The development of science models within \fink\ is completely driven by the user community\footnote{Users can propose science modules following instructions at \url{https://fink-broker.org/joining/}}, who is responsible for development and validation of its outputs.} Each team responsible for a classifier typically provides a pre-trained model, as well as the snippet of code necessary to run the inference on one alert\footnote{The {\sc Fink} engineer team provides examples to manipulate alert data, and all code and models can be found online at \url{https://github.com/astrolabsoftware/fink-science}.}, and the {\sc Fink} engineering team integrates it into {\sc Fink} for streaming processing at scale. 

Given the large volume of data, we developed a new service for the community during the challenge, the {\sc Fink} Data Transfer service.\footnote{\url{https://fink-portal.org/download}}
The {\sc ELAsTiCC} training set was made available via this service, which enables easy distribution of large volumes of data for many decentralised users. It also allows users to select any observing nights, apply selection cuts based on alerts content, define the content of the output, and stream data directly anywhere. Since the start of the challenge, more than one billion alerts have been streamed via this service.

For the experiments described in this work, nine classifiers were deployed (see Sect. \ref{sec:models}). Some of the classifiers used in the challenge are also classifiers used to process the ZTF alert stream. The differences in data rate, schema and the data itself (available filter bands, cadence and  magnitude limit, among others) between the ZTF stream and the simulated ELAsTiCC data made the transition less easy than we had originally anticipated. From this point of view, the lessons learned from the ELAsTiCC challenge were paramount in preparing for the arrival of the LSST alert stream.

Throughout the classifier design phase and the challenge itself, we monitor classifier performance in terms of throughput (alerts processed/second/core) and memory usage (MB/core). LSST will impose stringent requirements on throughput (we expect a continuous flow of 10,000 alerts per exposure, i.e. every 30 seconds), while our computing infrastructure imposes constraints on memory usage (cores with 2 GB RAM each). As an example, while processing in real-time
the first version of the challenge (using 24 cores in parallel), 82.2\% of alerts were classified in less than 30 seconds (the time between when an alert enters {\sc Fink} and when it is fully classified by the nine classifiers), 90.5\% in less than a minute, and over 99.9\% in less than 10 minutes. Delays larger than expected are partly due to processing (classifier versions and performance have evolved over time), but the high values are mainly explained by human intervention in the computing infrastructure, which interrupted operations while we were processing live data. These interventions are not expected during normal LSST operations. Regular measurements of Fink operations performance (profiling) are analysed to check that requirements are met.
%%%%%%%%%%%%%%%%%%%%%%%%%%%%%%%%%%%%%%%%
\section{Metrics}
\label{sec:metric}
In a classification task, several metrics can be used to assess the performance of the classifier, such as the receiver operating characteristic (ROC) and precision-recall curves and the confusion matrix. These are built from the precision (P), recall (R, also called the true positive rate, or TPR), and false positive rate (FPR), which in a binary classification are defined as

\begin{align}
    \hspace{1cm}\mathrm{P} &= \frac{TP}{TP+FP}, \nonumber \\
    \hspace{1cm}\mathrm{R} &= \frac{TP}{TP+FN}\nonumber \\
    \hspace{1cm}\mathrm{FPR} &= \frac{FP}{FP+TN}, \label{eq:metrics}
\end{align}

\noindent with TP(N) the number of true positives (negatives) and FP(N) the number of false positives (negatives). Precision can be understood as the purity of the predictions, while Recall is its completeness or efficiency, and the FPR is the ratio of wrongly classified objects of the negative class (also known as the false alarm rate). The output of a binary classifier is a the probability of a light curve belonging to the class of interest. Thus, the quantities on Eq. \ref{eq:metrics} will depend on a chosen probability threshold. By varying this threshold, one can obtain curves for recall versus FPR (ROC) and precision versus recall. The area under these curves (AUC) can be used as a metric to assess the performance of the classifier, where a perfect classifier would have an AUC of 1, and an AUC of $0.5$ for the ROC corresponds to a random classifier. 

The binary case can be straightforwardly extended to the multi-class case by using a one versus all approach, wherein the problem is split into a binary classification case per class, gathering every class except one as the negative class. In this case, there will be a ROC and a precision-recall curve for every single class. 
\par Anther popular metric for classification task is the confusion matrix, a square matrix showing the number (or percentage) of objects classified in every combination true/predicted class. Both metrics are used to access the efficiency of classifiers described in this work.

%%%%%%%%%%%%%%%%%%%%%%%%%%%%%%%%%%%%%%%%%%%%%%%%%%%
%%%%%%%%%%%%%%%%%%%%%%%%%%%%%%%%%%%%%%%%%%%%%%%%%%%

\section{Classifiers in Fink}
\label{sec:models}

\rev{For this work, we used 4 algorithms for classification. These are meant to represent the two possible applications of the broker infrastructure: broad classifiers using deep learning \revv{(DL)}, represented by CATS (Sect. \ref{subsec:cats}) and SuperNNova (Sect. \ref{subsec:snn}), and binary classifiers using feature extraction and tree-based algorithms, represented by the Early SNIa (Sect. \ref{subsec:earlyIa}) and the SLSN (Sect. \ref{subsec:PISN}) classifiers.}

\rev{The class specific classifiers (Sects. \ref{subsec:earlyIa} and \ref{subsec:PISN}) represent the most common scenarios in which a team interested in a specific class can profit from the broker infrastructure. As an example, we used a tool that incorporates physical assumptions about thermodynamic behaviour of the transient \citep{Rainbow} in combination with traditional ML algorithms. The same infrastructure can be used to incorporate other requests coming from science teams.}

\subsection{CATS}
\label{subsec:cats}

Recurrent neural networks (RNNs) are a class of models adapted to work with sequential data. They do so by constructing hidden states that carry information from the previous part of the sequence \citep{RNN_beginning, RNN_gentleintro}. One of the main problems found in training RNNs was the vanishing gradients: When the input sequence was long, the successive derivatives during backpropagation tended to erase the gradient \citep{bengio_rnngradient}, which then cause later time steps to be `disconnected' from earlier ones (i.e. a low memory capacity). Long-short term memory units \citep[LSTM;][]{lstm}  were designed to avoid the vanishing gradient problem by keeping not only a hidden state but also a memory state across all time steps. By using gates, the network can learn what information needs to be kept, removed, or inserted to the memory vector, increasing the RNN's memory capacity. 

\begin{figure}
    \centering
    \includegraphics[width=\columnwidth]{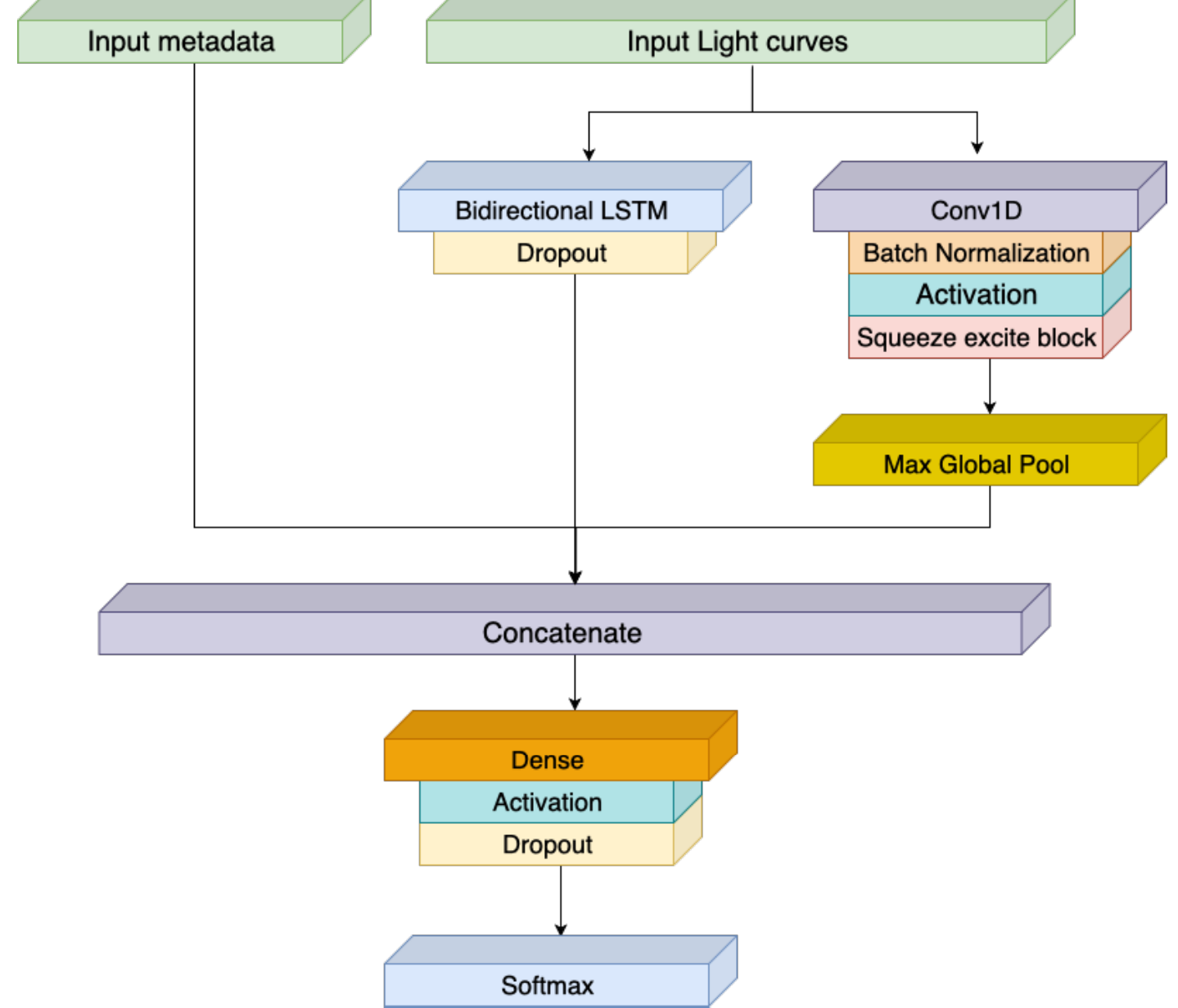}
    \caption{Illustration of the CATS architecture.}
    \label{fig:cats}
\end{figure}

The CBPF Alert Transient Search (CATS) was built by starting with a base network very similar to the Multivariate LSTM Fully Convolutional Network \citep[MLSTM-FCN;][]{mlstm-fcn}, using squeeze and excitation blocks; we use bidirectional LSTM layers and a series of fully connected layers before the output, adding a dropout layer after each of those. This architecture was shown to perform very well in several different time series tasks\footnote{See \url{https://paperswithcode.com/task/time-series-classification} for benchmarks on several different time series tasks.}, and the base architecture is shown in Fig \ref{fig:cats}. We then performed a hyperparameter search using keras-tuner, searching for the best configuration of number of convolutional, LSTM and Dense blocks, convolutional filters, LSTM and Dense layers' units and activation functions. This search was done using a subset of $10\%$ of the unique objects in our training sample to speed up the process. The final architecture consisted of one convolutional block with 32 filters, two bidirectional LSTM layers with 400 and 500 units respectively, and two fully connected layers with 64 and 96 units. All convolutional and fully connected layers are followed by a Rectified Linear Unit (ReLU) activation. 

Our inputs were the normalised flux and errors, (where both are normalised per light curve), \rev{and the difference in days from  the first available light curve data point  for that transient and the current alert detection}, using forced photometry when available. To that, the filter was added as an integer in the range $[1,6]$ corresponding to the LSST passbands, [$u$, $g$, $r$, $i$, $z$, $Y$]. All inputs were right-padded to match the longest light curve, so that the shape of the input is $(395, 4)$ (top-right box in Fig. \ref{fig:cats}). To this, we concatenated the Milky Way extinction (\texttt{mwebv}), host galaxy redshift (\texttt{hostgal\_zphot}), and the redshift of the transient (\texttt{z\_final}), when available, plus their errors (Input metadata) before passing the result to the fully connected layers. Using these features gave the best results when testing with a smaller subset of the data, slightly better than using only extinction and redshift of the host. 

We performed a K-fold cross-validation process to assess the robustness of our model. It consists of splitting the data into $k$ groups (folds), and performing $k$ iterations of training; at each iteration, one of the folds is used for validation, while the other $k-1$ are used for training. 
\par We used five folds; the unique object identifier (\texttt{diaObjectId}) was the parameter used to split the data, thus making every light curve of a single object either part of the training or validation set, with no contamination between them.
\par The model was trained for 15 epochs at each fold, with the \rev{NAdam optimizer \citep{nadam}}, on a cluster with eight NVIDIA RTX A6000. At each fold, the model where the validation loss was lowest was selected to perform the predictions. The architecture and training were implemented using TensorFlow 2 \citep{tensorflow}.
\par Table \ref{tab:cats_kfold} shows the metrics derived from the validation sets of the k-fold. The model performed satisfactorily, having a mean ROC AUC above $0.95$ for all classes, and a mean precision-recall AUC above $0.8$ except for the Long class, with little variance between the folds. Although the mean Precision for the Long class is approximately $80\%$, the Recall is less than $50\%$, due to half of the Long alerts being classified as SN-like events. The major type behind this confusion are Super luminous Supernovae (SLSN), which could resemble some types of supernovae when looking at short enough light curves. On the other hand, the $20\%$ rate of false positives of the Long class is dominated by SNII, which is also the dominant subclass; however, contamination by SNIa (the second most dominant subclass) is negligible. SNII with an extended plateau can be very similar to SLSNe, while SNIa are generally brighter and exhibit a characteristic feature in redder passbands, thus rendering their classification less controversial. Despite such caveats, results shown in Table \ref{tab:cats_kfold} confirm that CATS will be able to provide reliable classification results even under challenging data scenarios. 

\begin{table}[ht]
\caption{CATS results for the validation sample across all folds.}
    \centering
    \begin{tabular}{p{1.2cm}p{1.3cm}p{1.3cm}p{1.3cm}p{1.3cm}}
         & ROC & PR & \multirow{2}{*}{Precision} & \multirow{2}{*}{Recall} \\
         & AUC & AUC   &           &        \\
         \hline
    SN-like     & $0.99$ $({0.0001})$ & $0.99$ $({0.0002})$ & $0.97$ $(0.0022)$ & $0.99$ $(0.0013)$ \Tstrut\\
    Fast & $0.99$ $({0.0013})$ & ${0.80}$ $({0.047})$ & ${0.85}$ $({0.092})$ & $0.72$ $(0.017)$ \\
    Long & $0.96$ $({0.0012})$ & ${0.68}$ $({0.0041})$ & ${0.82}$ $(0.032)$ & $0.47$ $(]{0.041})$ \\
    Periodic & $1.00$ $(0.00001)$ & $1.00$ $(0.00003)$ & $1.00$ $(0.0002)$ & $1.00$ $(0.0001)$ \\
    Non-Periodic & $1.00$ $(0.00003)$ & $1.00$ $(0.0005)$ & $0.97$ $(0.003)$ & $0.96$ $(0.005)$
    \end{tabular}
    
    \tablefoot{Reported are the means with standard deviations in parentheses. ROC AUC and PR AUC are the area under the curve for the ROC and precision-recall curves, respectively, while the values for the precision and recall are taken from the confusion matrix averaged over all folds.}
    \label{tab:cats_kfold}
\end{table}

\subsection{SuperNNova}
\label{subsec:snn}

{\sc SuperNNova} \citep[{\sc SNN};][]{Moller:2020} is a DL light curve classification framework based on RNNs. {\sc SNN} makes use of fluxes over different band-passes and their measurement uncertainties over time for classification of time-domain candidates in different classes. Additional information such as host-galaxy redshifts and Milky Way extinction and their errors can be included to improve performance. 

{\sc SuperNNova} includes different classification algorithms, such as LSTM RNNs and two approximations for Bayesian Neural Networks (BNNs). Here, we only use the LSTM architecture, which was also used for classification of type Ia SNe (SNe Ia) in the Dark Energy Survey \citep{Moller:2022, Moller:2024}. For work on BNNs in the context of \textit{Rubin} see \cite{Moller:2022ICML}. 

We used the LSTM RNN to process the photometric time series and produce a sequence of hidden states. The sequence is condensed to fixed length through mean pooling. Finally, a linear projection layer was applied to obtain an N- dimensional vector, where N is the number of classes. A softmax function was used to obtain probabilities for an input to belong to a given class. During training, we randomly truncated light curves to improve the robustness of the classifier with partial light curves. {\sc SNN} is trained to optimize accuracy of balanced classes. 

We grouped observations in each passband within a given night. If a given filter is not observed, we assigned it a special value to indicate that it is missing. To deal with irregular time sampling, we added a delta time feature to indicate how much has elapsed since the last observation. We used the default configuration of {\sc SNN} with the normalisation as in \citep{Moller:2022} and added redshift and Milky Way extinction as additional features for this work.

Classification probabilities from {\sc SNN} can be used to select a sample by performing a threshold cut or by weighting the contribution of candidates by their classification score \citep{Vincenzi:2022,Vincenzi:2024, DES:5yr}. In this work we evaluated only the selection of samples using a probability cut set to $p>0.5$ or in multi-class classification, the largest probability for all classes.

We trained binary and broad class models. For the binary classification, we balanced the training set light curve numbers between the target class and other classes (randomly sampled). For the broad classifier we did also a balanced training set. However, the fast, long and non-periodic classes have considerably smaller numbers than the SN and periodic classes as shown in Fig.~\ref{fig:elasticc_train_broadclass_dist}; the balanced training set for broad classification was $\approx 2,000$ events per class.

We split the data set in $80\%$ for training, $10\%$ for validation and $10\%$ for performance evaluation. In Table~\ref{tab:snn_training}, we show the performance metrics for the different models obtained for this test set. 

For the broad model, we find that the \rev{lowest accuracy} classes are SNe and Long events. This may be due to the time-range of the light curve provided for classification as some Long events such as PISN and SLSNe during a reduced time range may resemble shorter time-scale SNe. We also find small confusion between Fast and Periodic transients that may be due to the same effect.

For the binary models, we also find that the model targeting Long events has lower accuracies than the other binary classification models. This may be due to the small data set used for training, which is composed of only thousands of light curves.

As expected, small and unbalanced training sets impact the performance of this framework, which was built for accurate classification with large and representative training sets. Further discussion on the performance of {\sc SuperNNova} with respect to training set size can be found in \cite{Moller:2020}.

\begin{table}
\caption{{\sc SuperNNova} results for a blind test sample.}
    \centering
    \begin{tabular}{llrrrr}
\multirow{2}{*}{Class} &  \multirow{2}{*}{Accuracy} &  ROC&  \multirow{2}{*}{Precision} &  \multirow{2}{*}{Recall} \\
 & & AUC & & \\
\hline
SN-like &        95.58 & 0.9890 & 96.19 & 93.38 \\
Fast &         98.15 &  0.9958  & 99.98  & 96.3 \\
 Long &        84.58 & 0.9270 & 83.85&  87.76 \\
Periodic &         99.93  & 0.9999  & 99.98  & 99.9\\
Non-Periodic &  99.4 &  0.9999  & 99.98  & 99.07 \\
\hline
Broad &         88.52 & - &  68.15 & 80.0\\ 
\end{tabular}
    \tablefoot{\revv{Results obtained on complete light curves using an independent test set from the training sample. All rows except the last one show the metrics for a binary target versus other types.}}
    \label{tab:snn_training}
\end{table}
%%%%%%

\subsection{Early supernova Ia}
\label{subsec:earlyIa}

Supernovae Ia (SNIa) were first used as standard candles in cosmological analysis in the end of the 20$^{\rm th}$ century, when they provided the first evidence of the Universe's current accelerated expansion \citep{riess1998, perlmutter1999}. Since then, large efforts have been devoted to the compilation of large SNIa samples, in the hope they can help unravel details about the behaviour of dark energy \citep[e.g.][]{aleo2023,Moller:2024}. 

Despite the undeniable impact large scale sky surveys can imprint on SN cosmology results,  such potential is strongly dependent on our ability to distinguish SNIa from other types of SN-candidates \citep[see, e.g.,][ and references therein]{ishida2019}. In the context of real data, labelling is an extremely expensive process and ideally we would like to discover such transients early enough so they are still sufficiently bright to allow spectroscopic classification. 

For ZTF processing, {\sc Fink} has an early supernova Ia classifier (hereafter, EarlySNIa) based on independent feature extraction for each of the two ZTF passbands and a random forest classifier enhanced by active learning \citep[for a complete description see ][]{leoni2022}. \rev{Its goal is to identify SNIa before or at peak, to optimally allow spectroscopic follow-up resources to be allocated}. The module has been successfully reporting EarlySNIa candidates to the Transient Name Server (TNS) since November/2020. However, in the context of LSST, with 3 times more passbands and a considerable sparser cadence, the module required significant modifications.

In order to allow classification with a lower number of points per filter and, at the same time, take into account colour information, we implemented the \textsc{Rainbow} multi-band feature extraction method proposed by \citet{Rainbow} to comply with the characteristics of the new data set. A parametric model was simultaneously fitted to the light curves in all available passbands, and the best-fit  parameter values were used as features, thus given as input to the random forest classifier \citep{ho1995random}. Assuming the transient can be approximated by a black-body, the framework combines temperature and bolometric light curve models to construct a 2D continuous surface. 
This approach enables early description even when the number of observations in each filter is significantly limited \citep[for more details see ][]{Rainbow}. 

\begin{figure}
    \centering
    \includegraphics[width=\columnwidth]{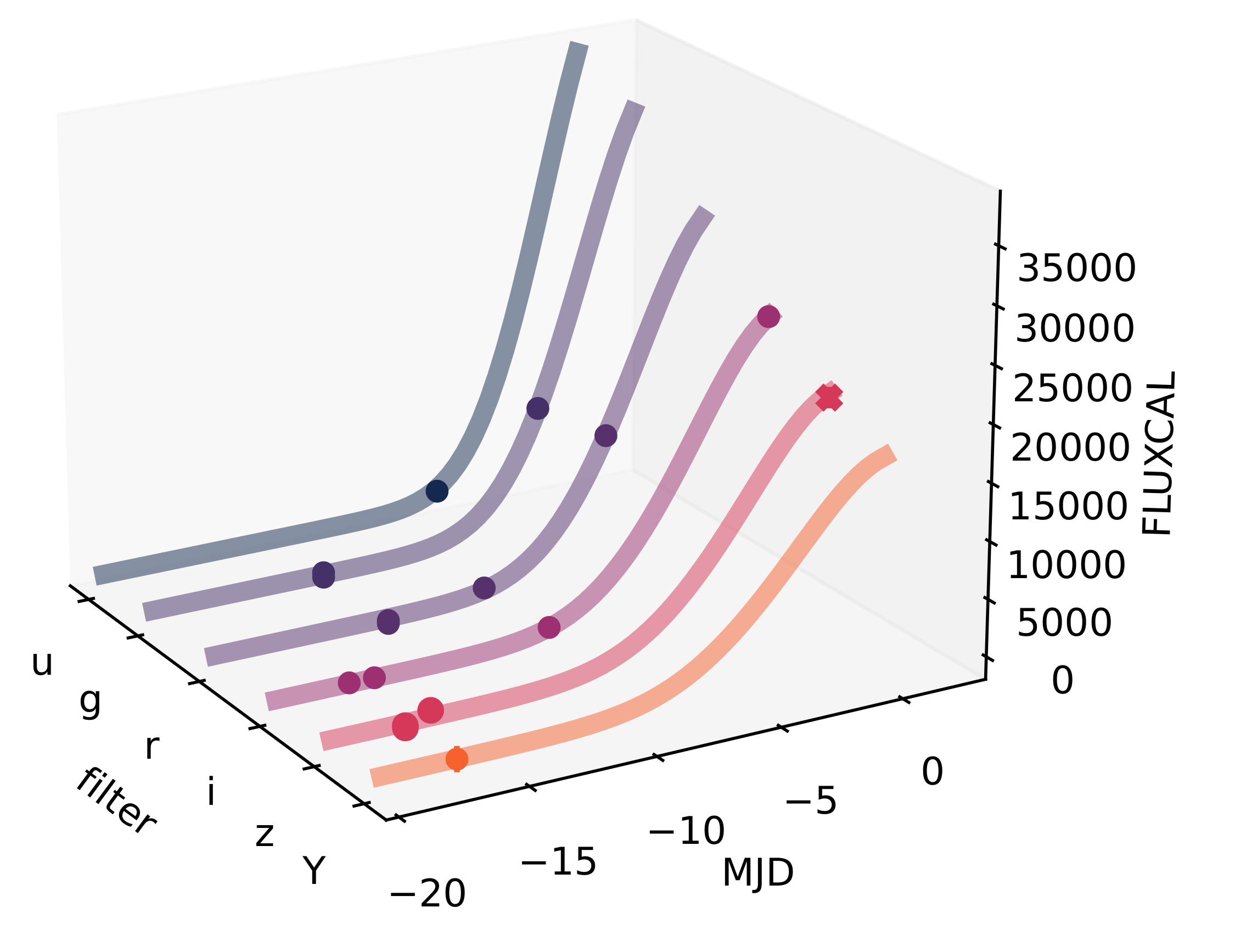}
    \caption{Example of light curve fit for an early SNIa performed with \textsc{Rainbow}. The estimated light curve behaviour in each filter (full lines) was found by using the photometric history within \texttt{alertId} = 244687224069 (circles). The most recent observation within that alert (cross) was added subsequently to illustrate the agreement between the estimation and measured value in the $z$-band.}
    \label{fig:earlySNIa_LC}
\end{figure}

The preprocessing for each alert included i) averaging all observations within the same night; ii) removing any intra-night flux measurements lower than -10 (\texttt{FLUXCAL} > -10); iii) requiring a minimum of seven points per object, in any filter, including forced photometry; and iv) ensuring that intra-night flux measurements are consistently increasing within at least two passbands. Thus, considering that only rising alerts survived such selection cuts, we described the bolometric evolution of our light curves with a logistic function of the form

\begin{equation}
f(t) = \frac{{\rm \texttt{amplitude}}}{ 1+ \exp( -\frac{t-t_{0}}{{\rm \texttt{rise\_time}}})} 
\label{eq:fit-sigmoid},
\end{equation}

\noindent where \texttt{rise\_time} is the characteristic time of rise, \texttt{amplitude} is the amplitude, and $t_0$ describes a reference time that corresponds to the time at half of the rising light curve. The temperature evolution was described with a falling logistic function of the form

\begin{equation}
T(t) = {\rm \texttt{Tmin}} + \frac{{\rm \texttt{delta\_T}}}{1 + \exp{\frac{t-t_{0}}{\rm \texttt{k\_sig}}}},
\label{eq:temp} 
\end{equation}

\noindent where \texttt{delta\_T} is the full amplitude of temperature, \texttt{Tmin} denotes the minimum temperature reached, \texttt{k\_sig} describes a characteristic timescale, and $t_{0}$ is a reference time parameter that corresponds to the time at half of the slope. We note that $t_{0}$ from the bolometric and temperature descriptions is used as a single common parameter whose role is to anchor functional behaviour, but it holds no physical meaning without a reference point, and thus it was not included in our final parameter set. Beyond these, \textsc{Rainbow} also returns a measurement of the quality of the fit (\texttt{reduced\_chi2}) and the maximum measured flux (\texttt{lc\_max}). Figure \ref{fig:earlySNIa_LC} illustrates the capability of the method in extrapolating the behaviour of a rising light curve even when the number of rising points is sparse. In this figure, the model (full lines) was fit using only the history within the alert (circles). The most recent observation (cross) was added subsequently as a way to compare the measurement with the prediction.  

As a result, each alert is represented by seven values. To this we added the mean signal-to-noise ratio (\texttt{FLUXCAL/FLUXCALERR}); the number of points in all passbands before intra-night smoothing\footnote{This number was obtained by counting the number of observations and each alert, not the corresponding metadata column.} (\texttt{nobs}); separation between the host and the transient (\texttt{hostgal\_snsep}) and the host photometric redshift (\texttt{host\_photoz}). Thus resulting in 11 parameters per alert. 

A total of \rev{5 334 911} alerts (\rev{791 779} objects) survived selection cuts, which corresponds to $\approx$ \rev{31}$\%$ of the \rev{alerts in the full}  training sample \rev{described in Sect. \ref{sec:elasticc}}. Among them, SNIa ($\approx$ \rev{34}$\%$),  SNII ($\approx$ \rev{28}$\%$), RRLyrae ($\approx$ \rev{11}$\%$), Ibc ($\approx$ \rev{8}$\%$) and \rev{AGN ($\approx$ 8\%)} were the most frequent classes. From these, we selected a sample of \rev{205 228 alerts (30\,000 objects)} \rev{for training a supervised ML model}. In this sample, the distribution of the major classes was unchanged. %For example, we ensured that the least represented class (KN) comprised $\approx 1\%$ of the sample used for training. 
We note that we report populations of individual classes to better illustrate the composition of the sample, but in \rev{reality}, we trained a binary classifier whose positive class corresponded to approximately a third of the full sample. \rev{The remaining 5 131 798 alerts (761 779 objects) were used for validating the classifier results. To avoid an information leak, we made sure to place all alerts from the same object either in the training or in the validation sample.}
%We used half of the unique object identifiers (\texttt{diaObjectId}) for training and the other half for validation to avoid information leak between the two subsets. 

We trained a random forest model using a scikit-learn \citep{scikit} implementation, using 50 trees, maximum depth of 15 and set the minimum number of alerts per leaf to be 0.01 \% of the training size. In the \rev{validation} sample, this resulted in \rev{P = 0.70/R = 0.70}, considering a probability threshold of 0.5.

%%%%%%%%%%%%%%%%%%%%%%%%%%%%%%%%%%
%----------------------------
%%%%%%%%%%%%%%%%%%%%%%%%%%%%%%%%%%%%

\subsection{Superluminous supernovae}
\label{subsec:PISN}

Superluminous supernovae (SLSN) are SNe whose peak optical luminosity
exceeds $-21$\,mag (see, e.g., \citealt{moriya2018} for a
review). Their rise times can vary between $\sim20$ days to more than 100\,days for some events, but their post-maximum decline rates are either consistent with $^{56}$Co decay (at least initially), or significantly
faster. This suggests that some SLSN can have a possible thermonuclear origin, which given the mass required can only be explained by population III star origin. At such mass, a pair production mechanism triggers and results in an instability that eventually leads to the collapse of the core, hence pair-instability SNe (PISN).
Currently, there have been no direct confirmation of their
existence, although good candidates have been reported  \citep{moriya2022, snad160}. The discovery and characterisation of a PISN would significantly impact our understanding of the connection between chemical evolution and structure formation in the Universe \citep{lsst2009}.
\par
\begin{figure}
    \centering
    \includegraphics[width=\columnwidth]{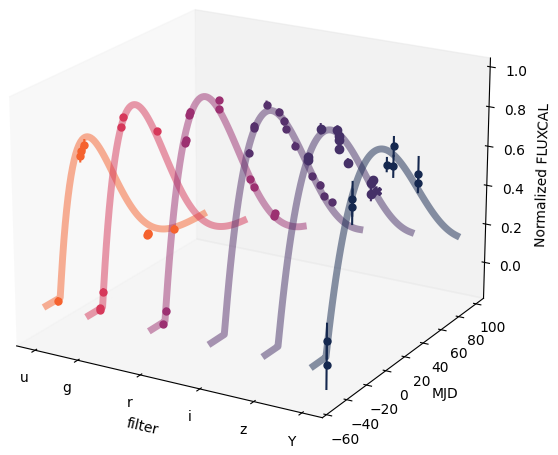}
    \caption{\rev{Example of a \textsc{rainbow} fit on an SLSN light curve (\texttt{diaSourceId} = 35707054076) using Eq. \ref{eq:SLSN} for the bolometric component and Eq. \ref{eq:temp} for the temperature component}. The flux is normalised \revv{to have a maximum at one}.}
    \label{fig:slsn}
\end{figure}

 Given that predicted light curves of SLSNe and PISNe can have a similar morphology, we use a common classifier for both and call it the SLSN classifier. Its core implementation is based on a feature extraction of normalised alerts followed by a random forest classification. For each filter we computed the following set of features: maximum and standard deviation of the flux; mean signal-to-noise ratio and number of points. We also added the following metadata information: right ascension (\texttt{ra}), declination (\texttt{dec}), host galaxy photometric redshift (\texttt{hostgal\_zphot}), host galaxy photometric redshift error (\texttt{hostgal\_zphoterr}) and distance between the host and the transient (\texttt{hostgal\_snsep}). In addition, parametric fits of the light curves are computed and best-fit values are used as parameters. Similarly to Sect. \ref{subsec:earlyIa}, we used the \textsc{Rainbow} \citep{Rainbow} framework to obtain a multi-passband description of the light curves. In this context, the bolometric flux is modelled using the following three-parameter function:

\begin{equation}
f(t) = \textit{max}\ (0, A(t - t_0) \times e^{-\frac{t - t_0}{t_{fall}}}),
\label{eq:SLSN}
\end{equation}

\noindent which depends on amplitude ($A$), a time offset ($t_0$) and a characteristic time of decay ($t_{fall}$). We found this simple functional form to be effective in classifying both SLSN-I and PISN. This equation was obtained by applying the Multi-View Symbolic Regression algorithm \citep{russeil_mvsr} to  real ZTF light curves from the SLSN candidate \texttt{SNAD160}\footnote{\url{https://ztf.snad.space/dr17/view/821207100004043}} \citep{aad_ztf}. \rev{The max operation has been manually added to prevent negative fluxes. We chose to model the temperature using Eq. \ref{eq:temp}, which enables the description of cooling transients}. An example of a SLSN fit is shown in Fig. \ref{fig:slsn}. It highlights the good agreement of the blackbody approximation to SLSN observations. The fit is performed using \textsc{light-curve}\footnote{https://github.com/light-curve/light-curve-python} python package, which internally uses iminuit minimising the negative log likelihood. The optimised parameters and the loss are included as features for the classifier. Thus we imposed that alerts contain at least seven observed points. In total, 26 features are extracted for each alert.\\

\par The classifier was based on a scikit-learn random forest algorithm trained using the active learning (AL) procedure proposed in \cite{leoni2022}. This strategy allows the classifier to focus on the relevant boundaries between SLSN and similar transients rather than simply learning the global distribution between  very unbalanced classes. This procedure tends to favour purity over completeness, which is reasonable given the volume of alerts that will be produced by LSST. To proceed we randomly sampled \rev{2} million alerts \rev{from the full training sample (Sect. \ref{sec:elasticc})}, which were further equally divided in training and validation samples. We queried 6 alerts at a time for 3500 loops (further queries resulted in overfitting, leading to performance decrease) for a final training sample composed of 21100 alerts. We set the maximum depth of trees and minimum number of alerts per leaf are set to be 15 and 0.01 \% of the training size, respectively. The model achieved \rev{90.8} \% purity and \rev{52.4} \% completeness \rev{in the validation sample}.
%%%
%%%
%%%
%%%
%%%
\section{Performance of classifiers}
\label{sec:class_perform}

In this section we evaluate the different classifiers available in {\sc Fink} using the test sample composed of all alerts from objects not in the training set of ELAsTiCCv1. In Sect.~\ref{sec:combined} we show how combining different classifiers can boost classification results. 
\par Figure \ref{fig:test_metrics} shows the confusion matrix for each classifier; each cell in the confusion matrix displays the \rev{results normalised by the predictions on top (bold), and by the true values at the bottom (italics).}  Besides these overall metrics, we analyse in Fig. \ref{fig:metrics_evolution} how metrics change with the number of detections and the true host redshift (\texttt{ZHELIO} from the truth table). This allows us to assess the ability of classifiers to generate unbiased samples, and their capabilities as a tool to select follow-up sources close to the time of detection.
\par \rev{Machine Learning metrics such as ROC and precision-recall curve can be found in Appendix \ref{ml_metrics}.}

\subsection{CATS broad classification} 
\label{sec:cats_results}
Figure \ref{fig:cats_metrics} shows the metrics for the CATS broad classifier. Similar to the cross-validation results presented in Sect. \ref{subsec:cats}, the model performs excellently with the SN-like, Periodic and Non-Periodic classes, having AUCs above $0.95$ for all of them. However, again similar to the earlier results, it had some issues with the Long \rev{and Short classes:} more than half of the Long alerts were classified as SN-like, \rev{and about $30\%$ of the Fast alerts were classified as Periodic, lowering the recall for both classes when compared to the other three. However, the purity for the Fast class is still above $90\%$, while for the Long class the value is $83\%$, significantly lower than for all others.} 

%\par The Fast class is where the results differ the most between the first and the last two years of simulations. The AUC of Precision-Recall curve dropped to $0.63$ compared to $0.82$ in first year, driven by a decrease in the number of Fast alerts correctly classified by our model. More than half of the Fast alerts were classified as Periodic, the same confusion the model had in the training set albeit to a smaller degree. A possible reason for this behaviour is the different distribution of number of detections for the two sets. Most correctly classified alerts of the Fast class in the training set have between 20 and 180 points. Since the training set is capped at one year, they form the majority of the total number of alerts, making the recall larger. On the test set, however, CATS incorrectly classified every alert with more than 200 points (there are very few light curves this size in the first year). Since it spans two years, a sizeable number of alerts have more than 200 points, contributing to a lower recall than in the training set. Due to this decrease in recall, the precision increased to above $90\%$, making CATS a useful tool to generate pure transient samples of every class. If we restrict the test set to only alerts with 150 points or less, the results are very similar to the training set, showing that the main contributing issue to the difference in performance is the difference in light curve length.
%%%%%%%%%%%%%%%

\par \rev{In Fig. \ref{fig:cats_sn_ndets}, it can be seen that CATS is able to correctly classify samples with less than ten detections, obtaining a precision of over $80\%$ for every class except Long, and nearly $100\%$ for SN-like, Periodic, and Non-periodic events. CATS is able to successfully classify these objects very early, an important quality when considering the schedule of follow-up observations. Also, Precision remains nearly constant as the number of detections grow for all classes (except for the Long class), showing the robustness of the model. }
\par \rev{Precision for the Long alerts increase as the number of detections grow since the model starts to better distinguish SN-like from SLSN alerts as the features from longer light curves begin to show. As expected from the results in Fig. \ref{fig:cats_metrics}, recall for both the Fast and Long classes increase as the number of detections grow, while SN-like and Periodic alerts have near-constant recall for all light curve lengths. Interestingly, very short Non-periodic light curves are classified as SN-like by the model, producing a low recall for Non-periodic alerts with less than five detections.}
%However, above $\sim 120$ detections, the precision for SN-like alerts decrease as the number of detections grow, due to SLSN alerts being classified as SN-like. This is understandable since one of the defining characteristics of SLSNe light curves is a longer decay time, which can make long SN look like SLSN especially when brightness is not a factor, since the light curves were normalised. Recall for the SN-like class behaves similarly, but the decrease only happens after $\sim 150$ detections, the model misclassifies SN-like alerts as non-periodic. A similar trend is seen for the Fast class, but now the confusion is with Periodic alerts. 
\par When looking at the metrics as a function of  redshift, the precision for SN-like alerts drop as the simulated redshift increases: this is due to SLSN alerts being classified as SN-like. The opposite is seen for the Long class, where low redshift alerts are classified as SN-like. \rev{Recall, on the other hand, remains almost constant for the full redshift range.}
%%-------------------------------------

\subsection{{\sc SuperNNova} as a broad classifier}

We show the performance metrics and confusion matrix in Fig.~\ref{fig:snn_broad_metrics} for {\sc SNN} broad classifier. As found when validating the model, both the SN and Long classes have large classification confusion. In Fig.~\ref{fig:snn_broad_ndets} we see this confusion is reduced with more detections as the classifier disentangles Long and SNe light curves with more precision. 

As with the CATS classifier, the Fast class is where the results differ the most between the \rev{training and test metrics} with a similar trend of confusion between Fast and Periodic light curves. 
% first and the last two years of simulations with a similar trend of confusion between Fast and Periodic light-curves. 

We highlight, that the loss used to train {\sc SuperNNova} is optimised for classification with large and representative training sets. In \cite{Moller:2020}, we have shown that $\approx 10^5$ light curves per class for training are necessary to achieve its top performance. To improve performance with {\sc ELAsTiCC}, the loss could be modified to allow the usage of non-balanced training sets and/or larger training sets could be constructed with additional simulations or augmentation techniques. We leave this task for future work.
%%%%
\subsection{{\sc SuperNNova} binary classifiers}

We find better performance for the SN-like classification with the binary classifier (Fig. \ref{fig:snn_bin_snlike_metrics}) than with the broad one \rev{in particular for recall}. This suggests, as expected, that the increase of the training set for the target is extremely important for our algorithm. We also find an improvement for classes with smaller training sets such as the Long class (Fig. \ref{fig:snn_bin_long_metrics}), \rev{although the model still has troubles with this class. The Fast class remains the hardest to classify with this algorithm (see Fig. \ref{fig:snn_bin_fast_metrics}).}

The evolution of metrics as a function of number of detections and redshift for the {\sc SNN} \rev{binary and broad classifiers is shown in Figs. \ref{fig:snn_bin_ndets} and \ref{fig:snn_broad_ndets}. The same trend is found for both the broad and binary classifiers, where the precision for the Non-periodic and Long classes increase as more detections are available.} \rev{Periodic and SN-like classification show high performance with low number of detections.} Thus, we expected a good performance in the classification of early light curves for these classes. This is an important feature when scheduling follow-up observations, as explored for \textit{Rubin} SNe Ia in \cite{Moller:2024}.

We highlight that the binary classifiers presented in this work were all trained in a similar manner without tuning {\sc SNN} to the different targets. Depending on the science goal, data curation, loss and algorithm hyper-parameters could be adjusted to improve its performance. For example, to improve Fast transients classification we could train with shorter light curves for all classes, an augmented training set or a modified loss to tackle small training sets. As shown in \cite{Moller:2022} for SNe Ia, an adequate light curve time span selection for a given goal to reduce non-transient detections improves performance.
%%%%%%

\subsection{Early supernova Ia}

\begin{figure*}
    \centering
    \begin{minipage}[t]{.32\textwidth}
    \centering
    \subfloat[CATS]{\includegraphics[width=\textwidth]{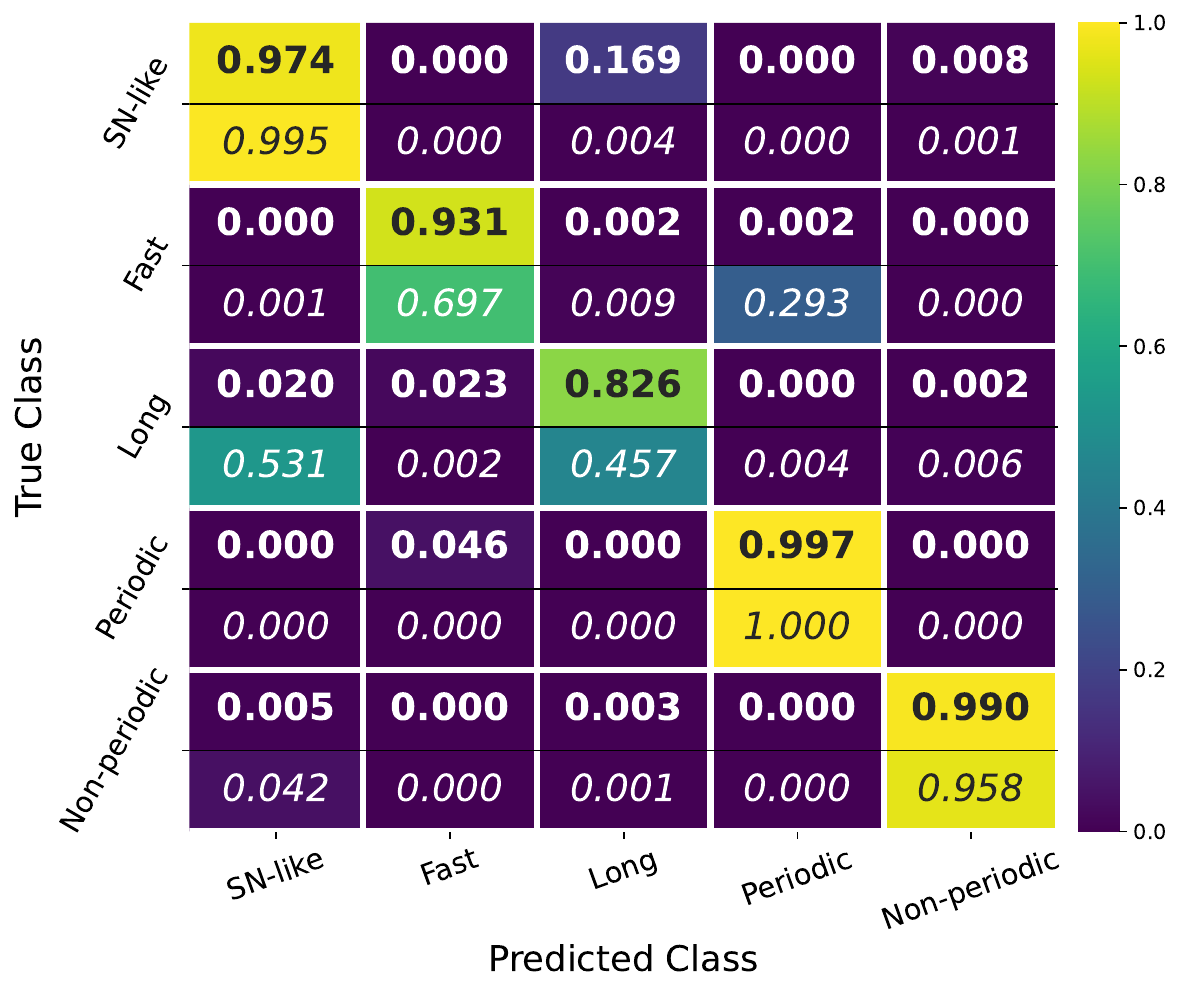}\label{fig:cats_metrics}}
    \end{minipage}
    \begin{minipage}[t]{.32\textwidth}
    \centering
    \subfloat[{\sc SuperNNova} broad]{\includegraphics[width=1.01\textwidth]{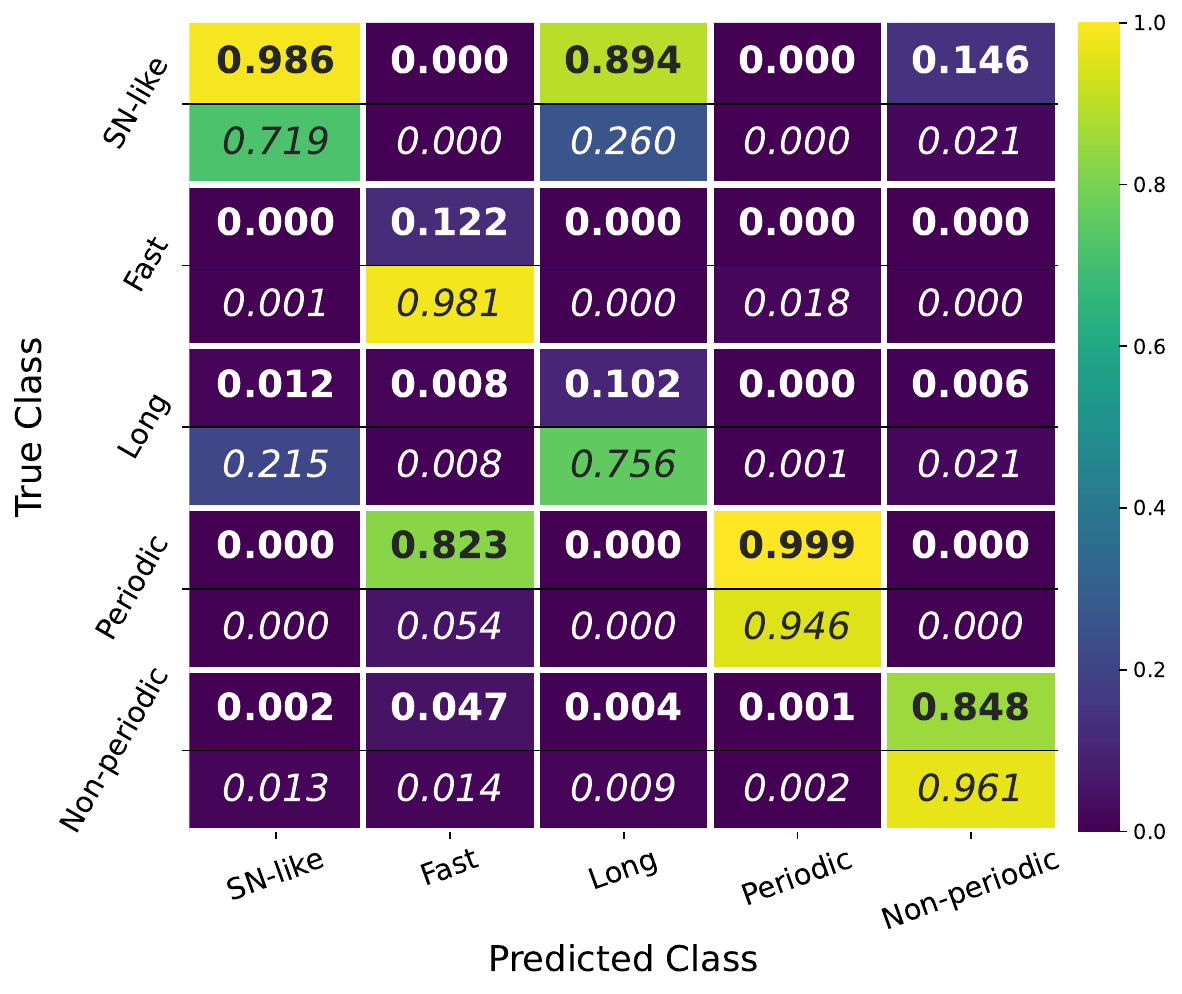}\label{fig:snn_broad_metrics}}
    \end{minipage}
    \begin{minipage}[t]{.32\textwidth}
    \subfloat[{\sc SuperNNova} binary for SN-like]{\includegraphics[width=\textwidth]{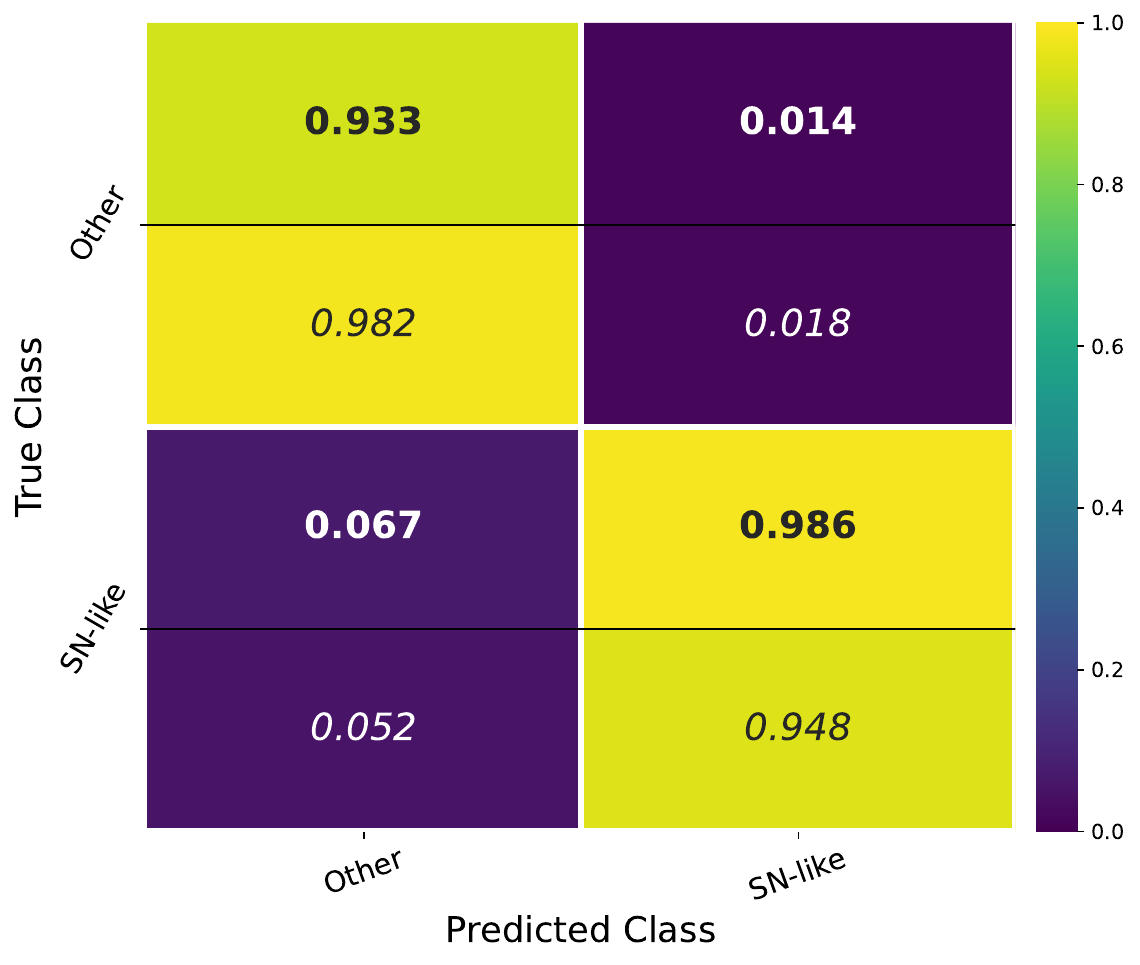}\label{fig:snn_bin_snlike_metrics}}
    \end{minipage}
    \begin{minipage}[t]{.32\textwidth}
    \centering
    \subfloat[{\sc SuperNNova} binary for Fast]{\includegraphics[width=\textwidth]{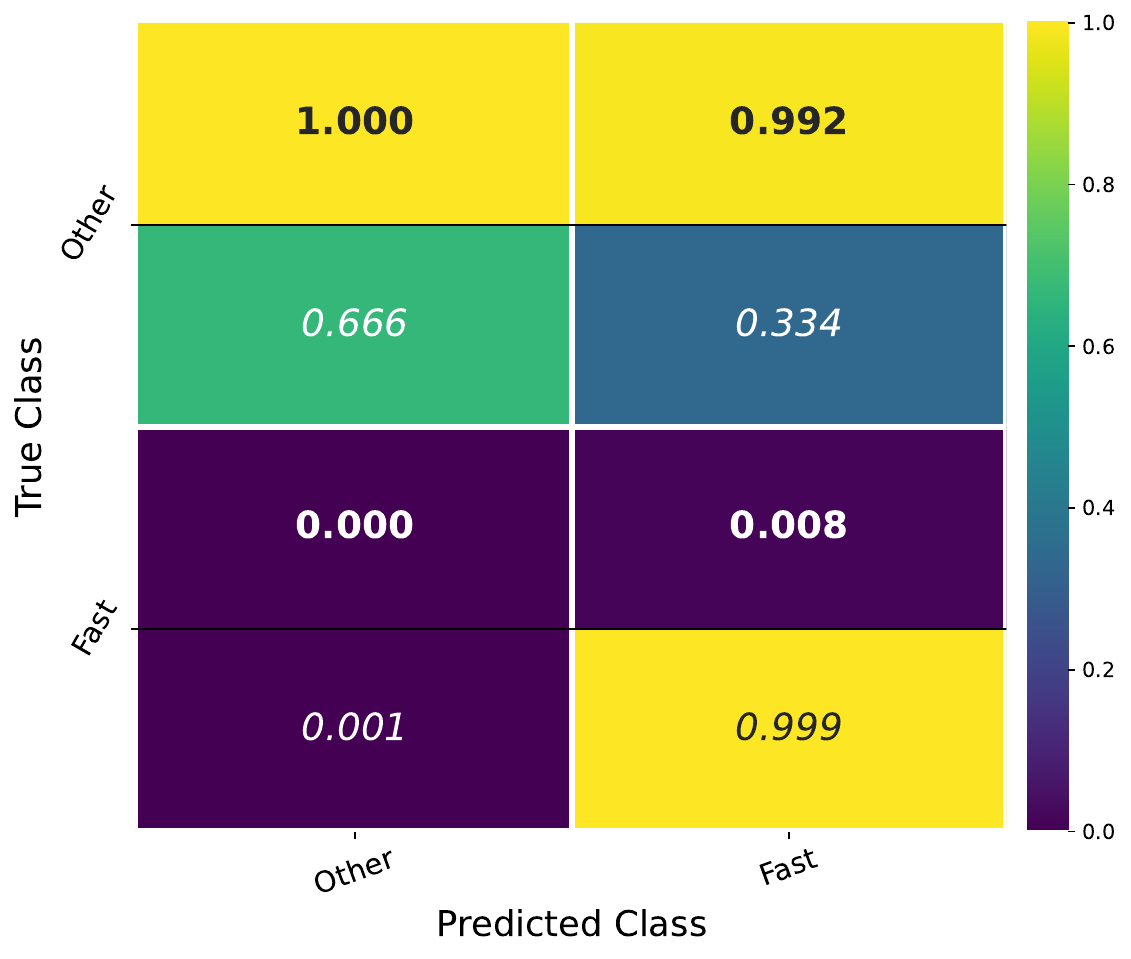}\label{fig:snn_bin_fast_metrics}}
     \end{minipage}
     \begin{minipage}[t]{.32\textwidth}
    \subfloat[{\sc SuperNNova} binary for Long]{\includegraphics[width=\textwidth]{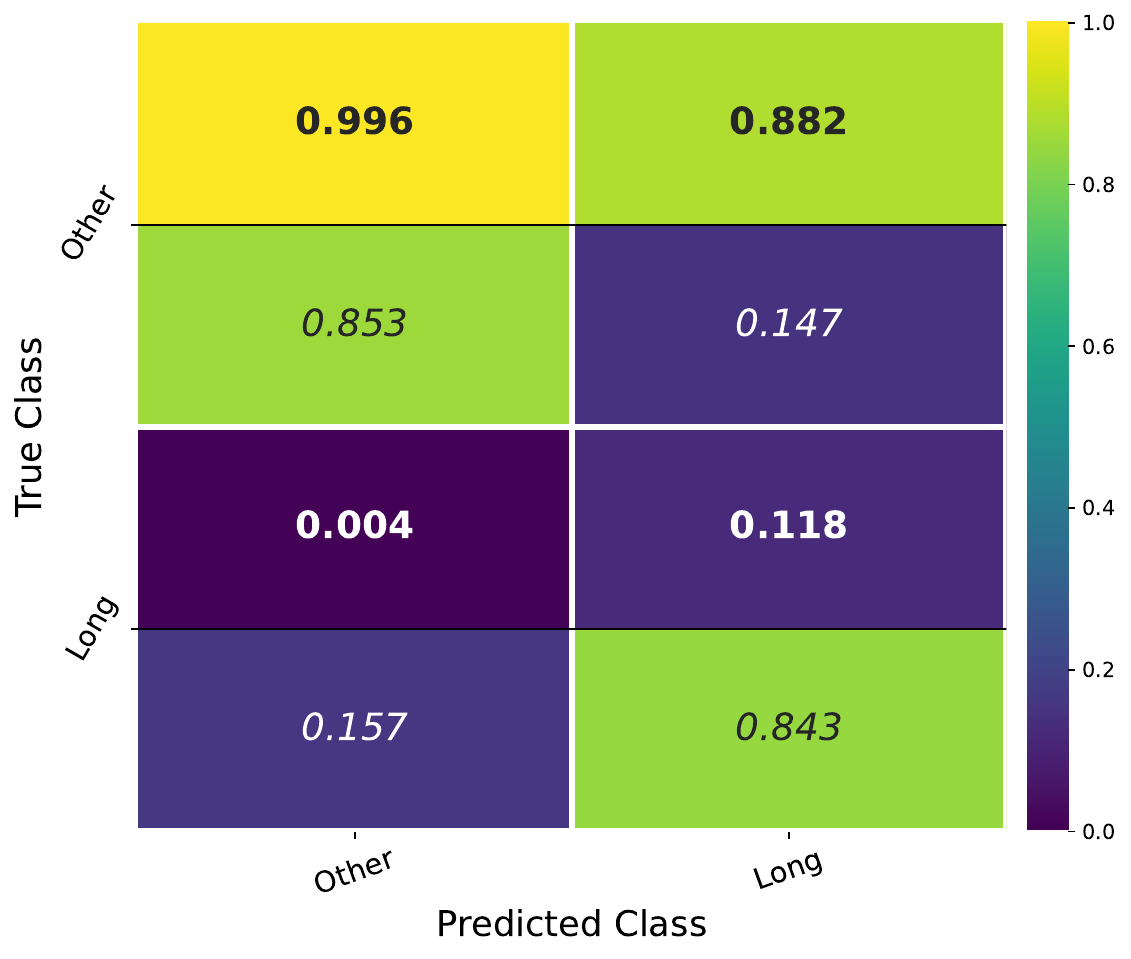}\label{fig:snn_bin_long_metrics}}
    \end{minipage}
    \begin{minipage}[t]{.32\textwidth}
    \centering
    \subfloat[{\sc SuperNNova} binary for Periodic]{\includegraphics[width=\textwidth]{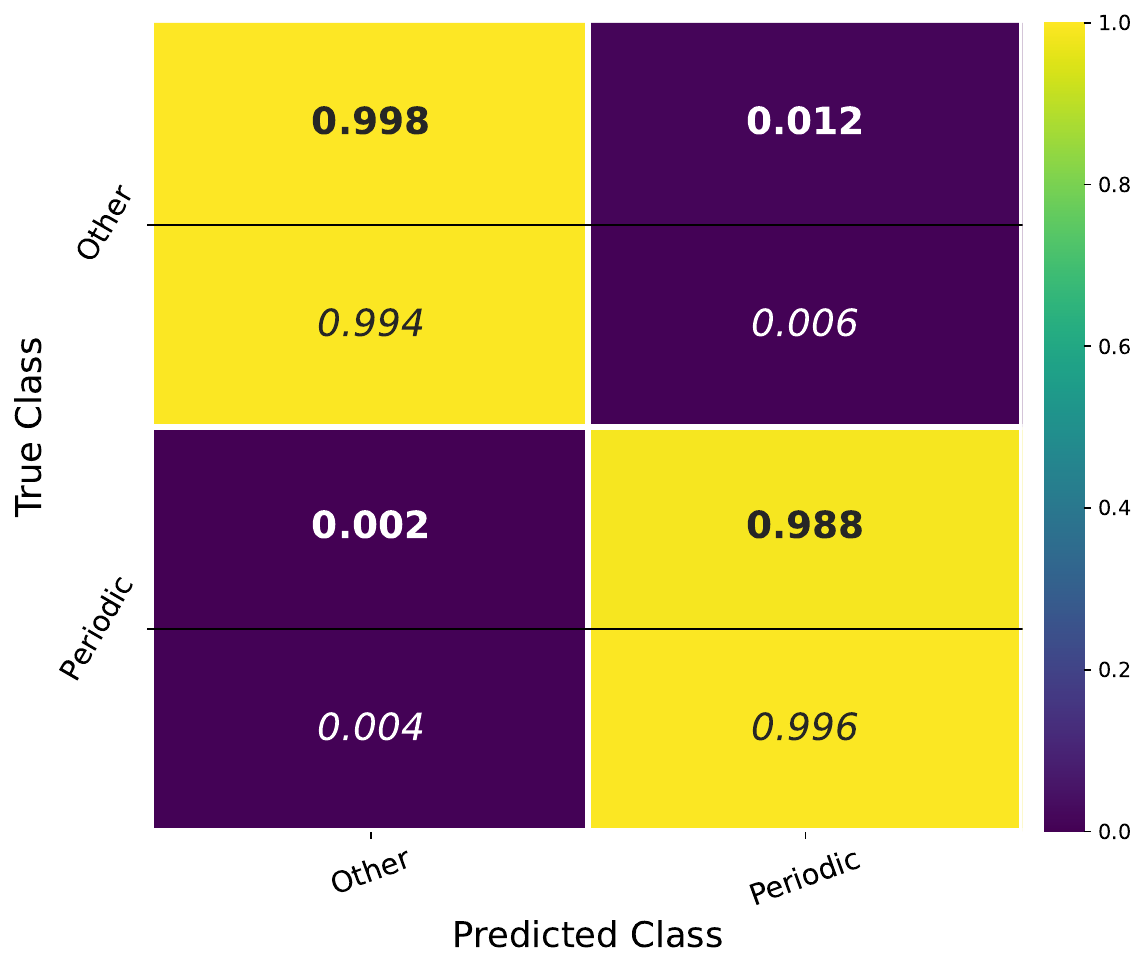}\label{fig:snn_bin_periodic_metrics}}
     \end{minipage}
     \begin{minipage}[t]{.32\textwidth}
    \centering
    \subfloat[{\sc SuperNNova} binary for Non-periodic]{\includegraphics[width=\textwidth]{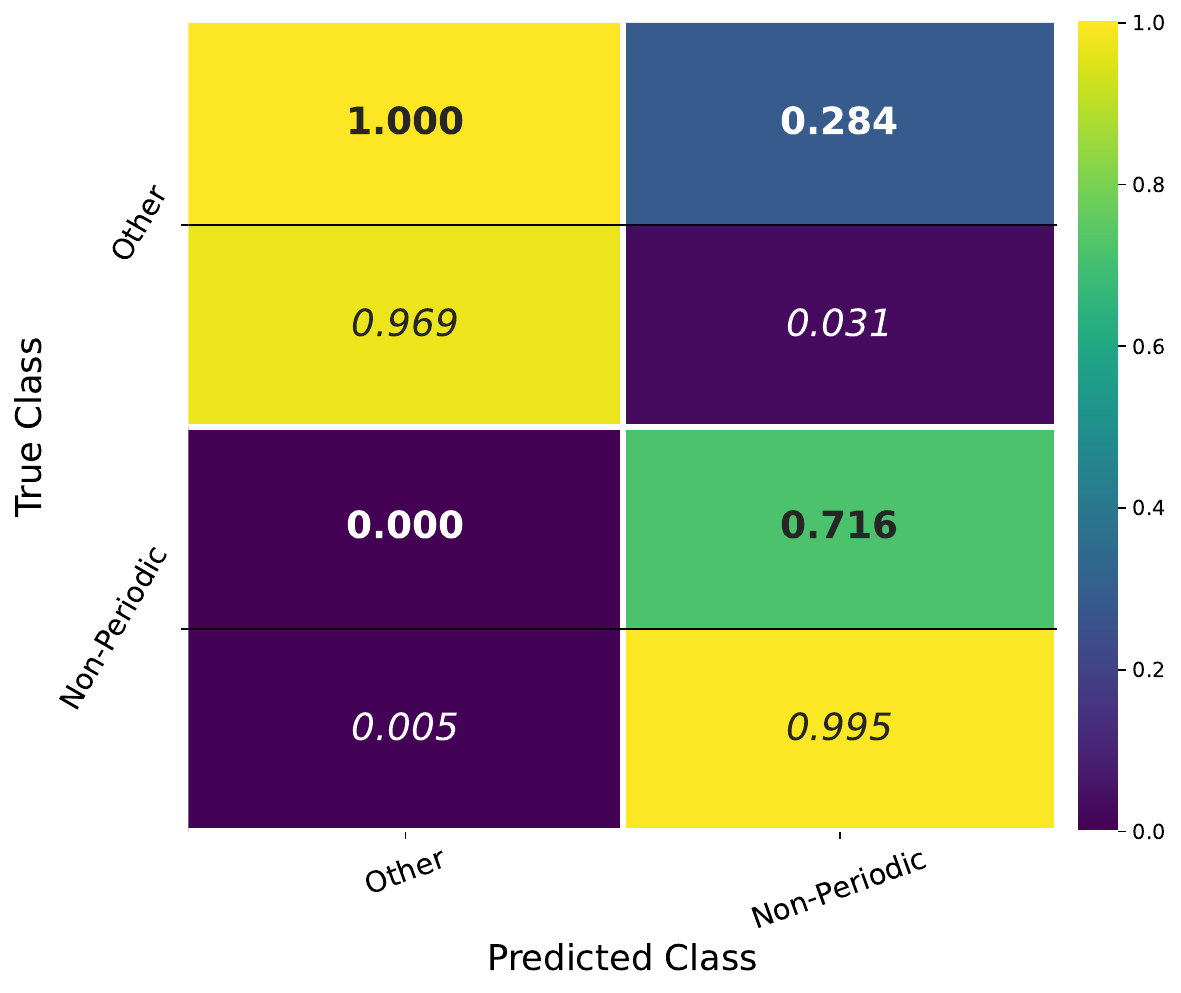}\label{fig:snn_bin_nonperiodic_metrics}}
     \end{minipage}
     \begin{minipage}[t]{.32\textwidth}
    \centering
    \subfloat[SLSN]{\includegraphics[width=\textwidth]{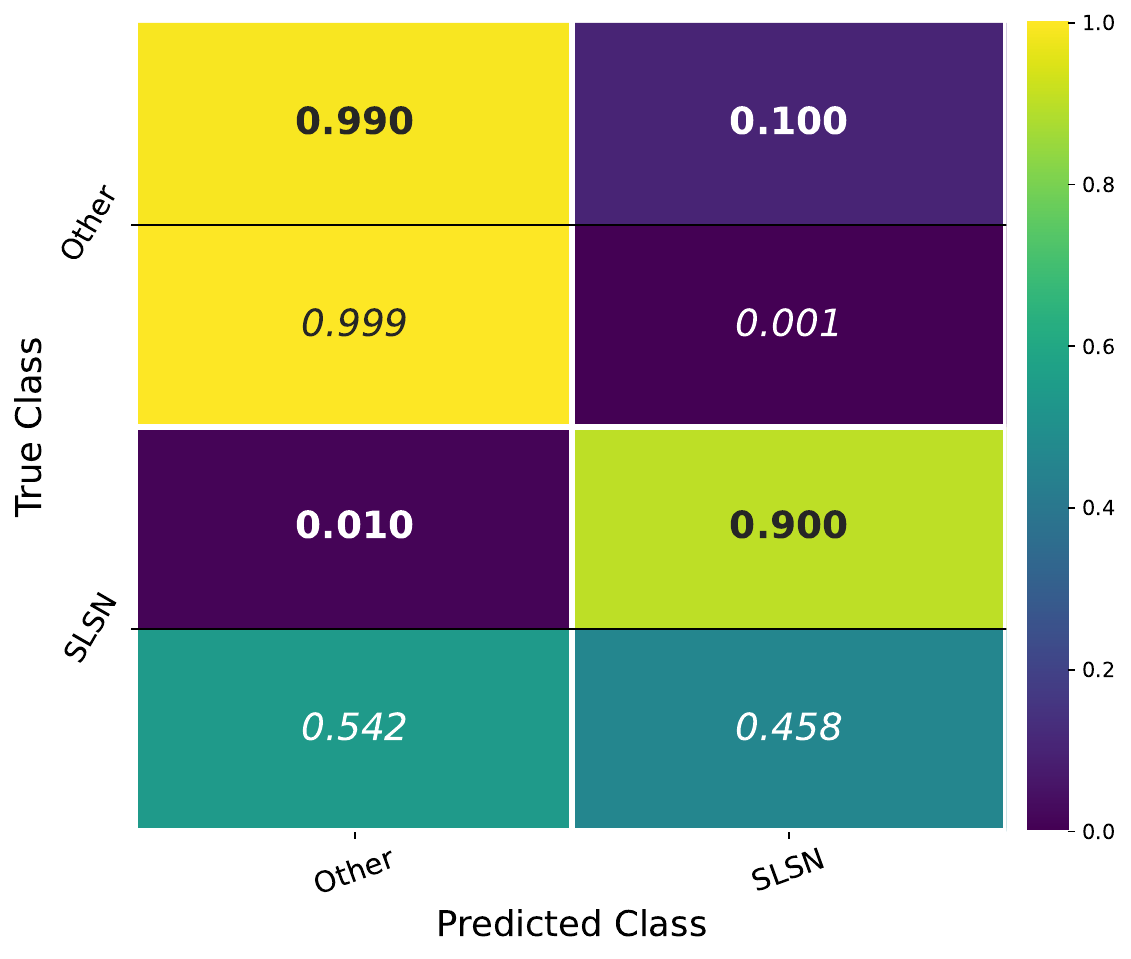}\label{fig:slsn_metrics}}
     \end{minipage}
     \begin{minipage}[t]{.32\textwidth}
    \centering
    \subfloat[Early SNIa]{\includegraphics[width=\textwidth]{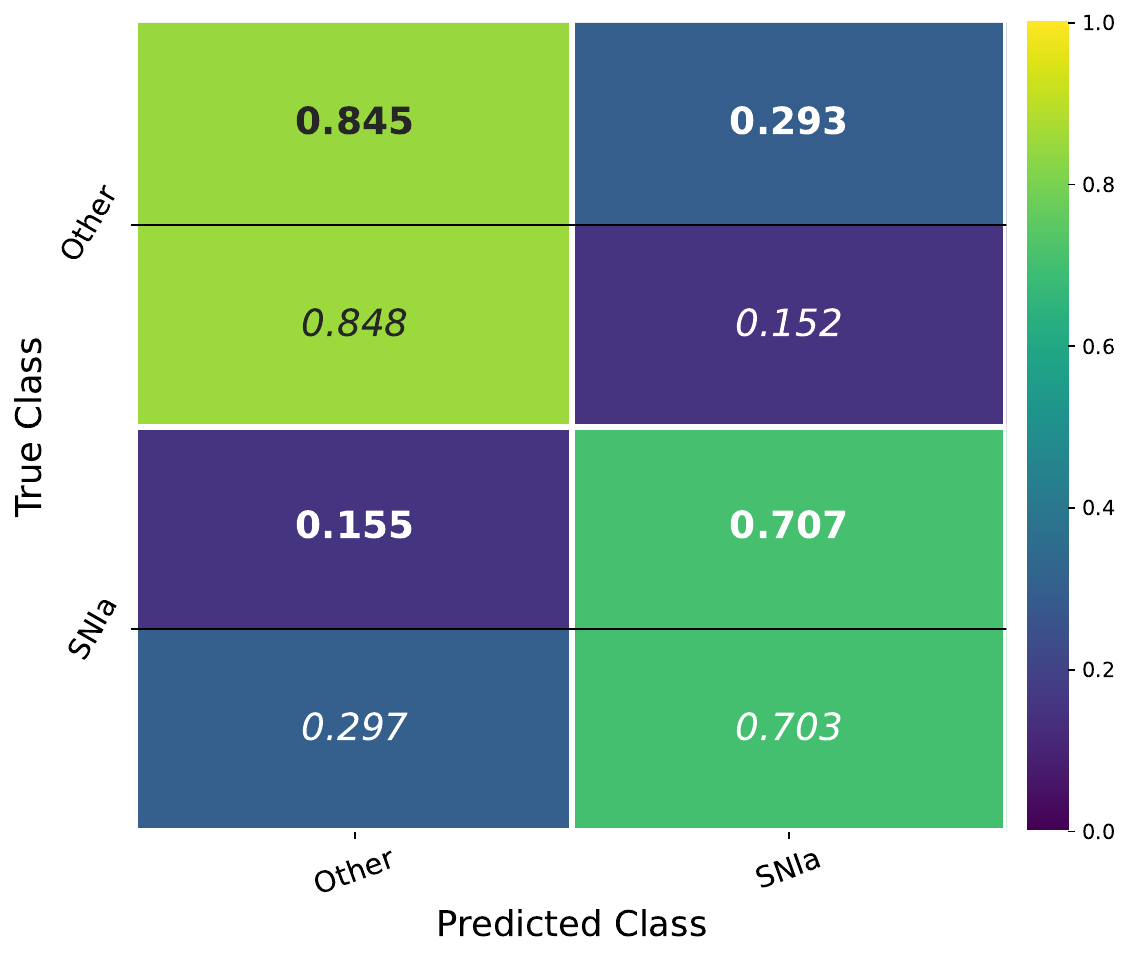}\label{fig:earlysnia_metrics}}
     \end{minipage}
    \caption{\rev{Confusion matrices for each classifier in the test test. For each cell in the confusion matrix, the top row (in bold) was normalised to the predictions and the bottom row (in italics) to the true values so that the main diagonal shows the precision in bold, on top, and recall in italics, at the bottom.}}
    \label{fig:test_metrics}
\end{figure*}

The performance of the EarlySNIa classifier is reported in Fig. \ref{fig:earlysnia_metrics}. One caveat we should keep in mind is that the module is only interested in classifying rising light curves. Thus, several alerts are eliminated by selection cuts, never being classified at all. Results presented here correspond to alerts that survived the feature selection (Sect. \ref{subsec:earlyIa}). Among these, the module was able to achieve $\sim0.7$ precision and recall. Meaning the precision was maintained while recall increased slightly when compared to results from the validation  sample. \rev{This matches the results found during training and represents $\approx 10\%$ decrease in comparison to those reported in \citet{leoni2022}, which reports results from the current EarlySN Ia module within \fink, applied to the ZTF stream. Considering the significant increase in volume and complexity present in the ELAsTiCC data set, we consider this a resilient and promising result. Once LSST start running, an active learning procedure similar to the one described in \citet{leoni2022} can be employed in real time, thus improving results found in this work}. The most common contaminants are: SN II ($\sim 18\%$) and SNIbc ($\sim 9\%$). All other classes correspond to the remaining 3\%, with SNIa-91bg and SNIax comprising $\sim 1\%$ each.

Figure \ref{fig:earlysnia_per_det} shows how classification results evolve with the number of detections and simulated redshift. We note that precision already starts higher than 0.6 for seven observed data points (the minimum requirement) and peaks around 20 photometric points, while recall remains almost stable even with more detections. The sample identified as EarlySNIa by the algorithm is highly skewed towards small light curves, with $\sim 75\%$ of them having ten detections or less. Moreover, since we are working only with rising behaviours, it is reasonable to expect that light curves with lower number of points will dominate the results\footnote{\rev{The long duration alerts are attributed to long history of forced photometry or variable sources with a significant burst in the last months of the survey. This also correlates with the steep decline in precision.}}. In this figure, we can also observe that classification results peak for redshift around 0.5 and degrades after that. This is expected since the SN is most likely to be discovered at or after maximum when at high redshifts.

%\begin{figure*}\ContinuedFloat
%    \centering
%    \begin{minipage}[t]{.48\textwidth}
%    \centering
%    \subfloat[{\sc SuperNNova} binary for Non-Periodic]{\includegraphics[width=0.95\textwidth]{figures/snnNonperiodic_both_cm.pdf}\label{fig:snn_bin_long_metrics}}
%    \end{minipage}
%
%    \begin{minipage}[t]{.48\textwidth}
%    \centering
%    \subfloat[SLSN]{\includegraphics[width=0.95\textwidth]{figures/slsn_both_cm.pdf}\label{fig:slsn_metrics}}
%    \end{minipage}
%
%
%    \begin{minipage}[t]{.48\textwidth}
%    \centering
%    \subfloat[Early SNIa]{\includegraphics[width=0.95\textwidth]{figures/earlySNIa_both_cm.pdf}\label{fig:earlysnia_metrics}}
%    \end{minipage}
%    \caption{Continued.}
%\end{figure*}

%\begin{figure*}\ContinuedFloat
%    \centering

%    \subfloat[EarlySNIa]{\includegraphics[width=0.95\textwidth]{figures/earlysnia_test_metrics_2.eps}\label{fig:earlysnia_metrics}}

%    \caption{Continued.}
%\end{figure*}

%%%%%%%%%%%%%%%%%%%%%%%%
\begin{figure*}
    \centering

    \subfloat[CATS]{\includegraphics[width=0.94\textwidth]{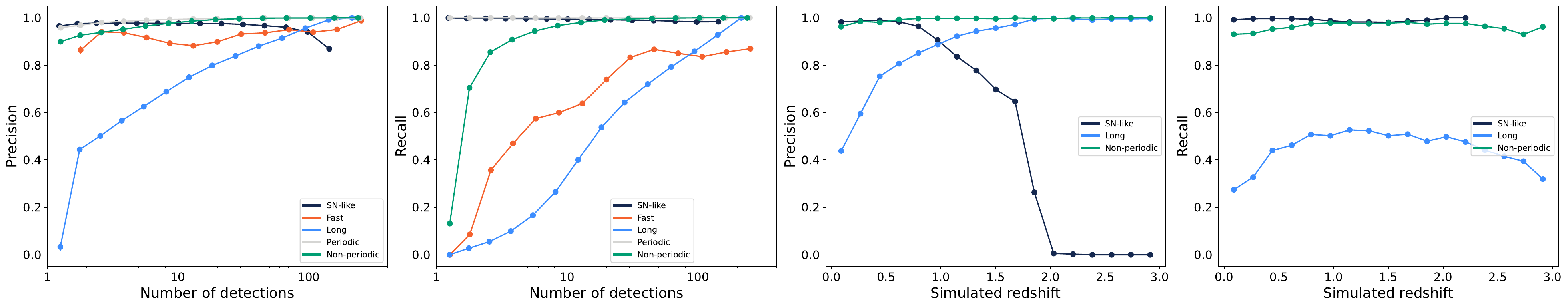}
    \label{fig:cats_sn_ndets}}

    \subfloat[{\sc SuperNNova} broad]{\includegraphics[width=0.94\textwidth]{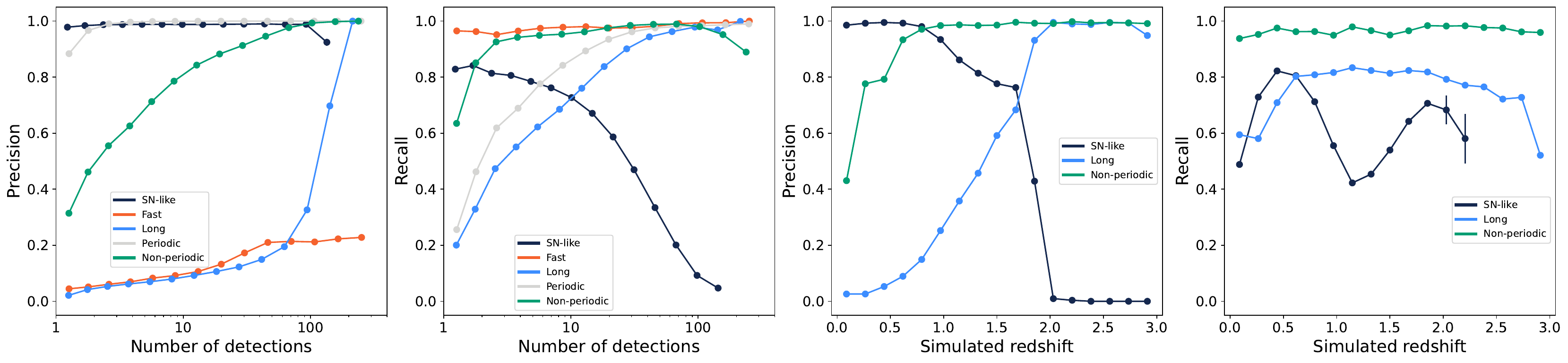}\label{fig:snn_broad_ndets}}

    \subfloat[{\sc SuperNNova} binary classifiers]{\includegraphics[width=0.94\textwidth]{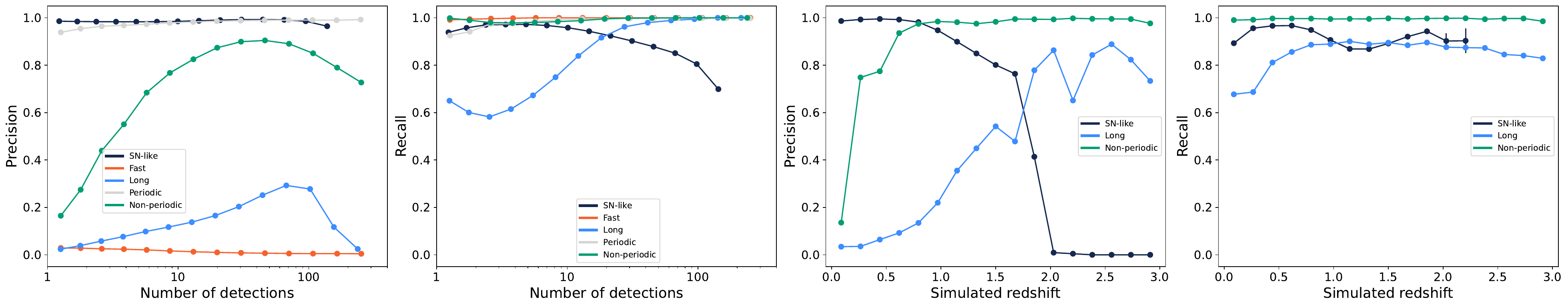}\label{fig:snn_bin_ndets}}

    %\subfloat[{\sc SuperNNova} binary for Fast]{\includegraphics[width=0.88\textwidth]{figures/snnFast_PR_ndet.pdf}\label{fig:snn_fast_ndets}}    

    \caption{Evolution of precision and recall as a function of the number of detections (left two panels) and host galaxy redshift (right two panels). Fast and periodic alerts have no redshift available and thus have only the first two panels.}
    \label{fig:metrics_evolution}
    
\end{figure*}

\begin{figure*}\ContinuedFloat
    \centering

    %\subfloat[{\sc SuperNNova} binary for Long]{\includegraphics[width=0.88\textwidth]{figures/snnLong_PR_ndet.pdf}\label{fig:snn_long_ndets}}

    %\subfloat[{\sc SuperNNova} binary for non-Periodic]{\includegraphics[width=0.88\textwidth]{figures/snnPeriodic_PR_ndet.pdf}\label{fig:snn_per_ndets}}

    %\subfloat[{\sc SuperNNova} binary for Periodic]{\includegraphics[width=0.88\textwidth]{figures/snnNonperiodic_PR_ndet.pdf}\label{fig:snn_nonper_ndets}}

    \subfloat[SLSN classifier]{\includegraphics[width=0.94\textwidth]{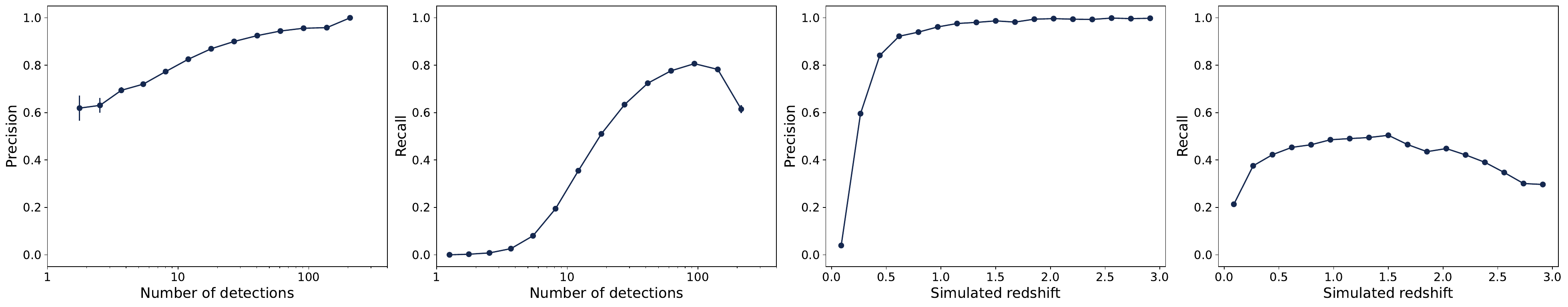}
    \label{fig:slsn_per_ndets}}

    \subfloat[EarlySNIa]{\includegraphics[width=0.94\textwidth]{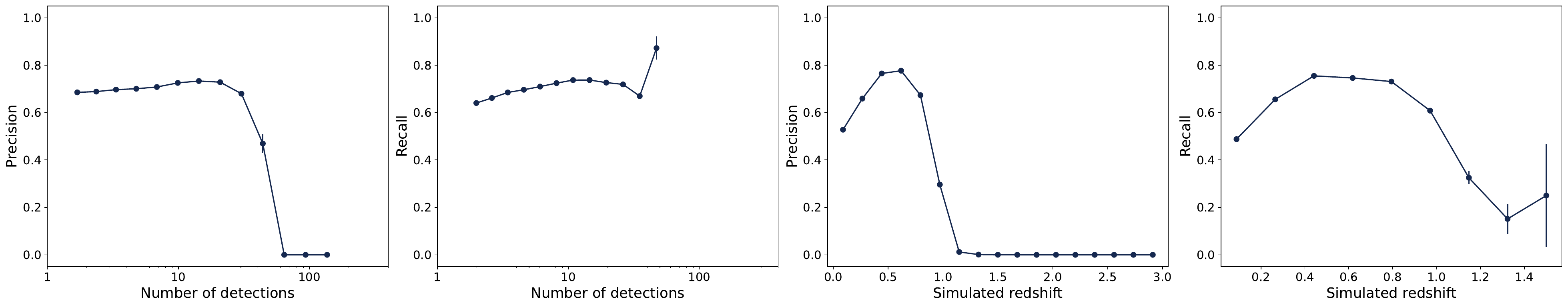}
    \label{fig:earlysnia_per_det}}
    \caption{Continued.}

\end{figure*}

%\begin{figure*}\ContinuedFloat
%    \centering
%    \subfloat[EarlySNIa]{\includegraphics[width=0.88\textwidth]{figures/earlysnIa_PR_ndet.pdf}
%    \label{fig:earlysnia_per_det}}
%    \caption{Continued}
%\end{figure*}

\subsection{Superluminous supernovae}
\label{subsec:SLSN_res}

Figure \ref{fig:slsn_metrics} displays the \rev{confusion matrix} of the binary SLSN classifier. It provides an excellent purity of \rev{90.0} \% and a recall of \rev{45.8} \%. In the context of the very large ELAsTiCC data set, we favour this high precision low recall asymmetry as it would still result in almost \rev{200K} SLSN alerts being classified with high confidence. \rev{However, as illustrated by Fig. \ref{fig:slsn_per_ndets}, the recall is highly dependent on the number of detections of the light curves. While most SLSN are missed when only a few data points are available, the recall increases to $\sim$50\% when 20 observations are available and up to $\sim$80\% for 100 detections. Purity-wise, the performance of the model linearly increases with the number of detections, ranging from $\sim$60\% to almost 100\%.} We observe that the classifier presents a clear under performance in precision for objects with redshift lower than  0.5. Indeed, SLSN are intrinsically bright objects that are more frequently found at high redshift (corresponding large observation volumes). Therefore, a large majority of the sample has a redshift above 0.5 (with a maximum around z=1). It results in a challenging learning task at low redshift,  which impacts the final performances. Outside this range, the classifier is conservative and therefore very reliable on positive answers. 

\rev{Although the purity of the classifier is very high, a study of the contamination reveals that almost two thirds of the alerts wrongly classified as SLSN are SNII. This result can be explained by the presence of SNIIn within the ELAsTiCC data set. This subtype of SNII, resulting from the interaction of the SNII with the circumstellar medium, is particularly bright and long-lasting. The most extreme ones even stand at the border between the SNII and the SLSN class, potentially forming a continuum between the two classes \cite{moriya2018}. Hence, the classifier contamination is to be expected and is particularly hard to reduce. }

\rev{Finally, a feature importance study has been conducted in order to characterize the decision process. The most important column for classification is the error on the fit, highlighting that the parametric model chosen is suited to describe SLSN events. The second and third most important features are parameters from the model, respectively the amplitude ($A$) and the minimum temperature ($T_{min}$). It once again demonstrates that the fit is key in the separation of the parameter space. The high relevance of the temperature parameter in particular indicates that the \textsc{Rainbow} framework enables the computation of informative features to distinguish transient events. }
%Compared to the testing sample, we observe a 6.6\% improvement of precision and a 17\% decrease in completeness. This variation in behaviour is explained by the long scale evolution of some SLSN, which may last more than a year. 
%Given that the training sample considered only the first year of ELAsTiCCv1 the model has not been trained to identify the longest SLSN. This is illustrated in Figure \ref{fig:slsn_per_ndets} where the peak of recall reaches up to 50\% between 50 and 150 detections, but largely drops with further detections where SLSN are missed. On the other hand, we observe that the precision linearly increases with the number of detections.

\section{Combining classifiers}
\label{sec:combined}

Given the wealth of developed classifiers targeting different science cases, combining some of them could provide better results. In this session we investigate the effectiveness of considering ensembles of two classifiers built from intrinsically different algorithms in order to boost classification results.

\subsection{Broad classifier as a first step}
\label{subsec:hierarchical}

We investigate the possibility of building a hierarchical classifier, where a broad classifier is initially applied to remove a large part of the contaminants, passing its results to a binary, more specific, model, possibly resulting in a more pure sample. We explore here using CATS before the SLSN classifier. Since the EarlySNIa binary classifier has the majority of contaminants within the SN-like broad class, a previous broad classifier is not expect to impact their results. 
%%%%%%%%%%%%%%
\begin{figure}[ht!]
    \centering
    \includegraphics[width=\columnwidth]{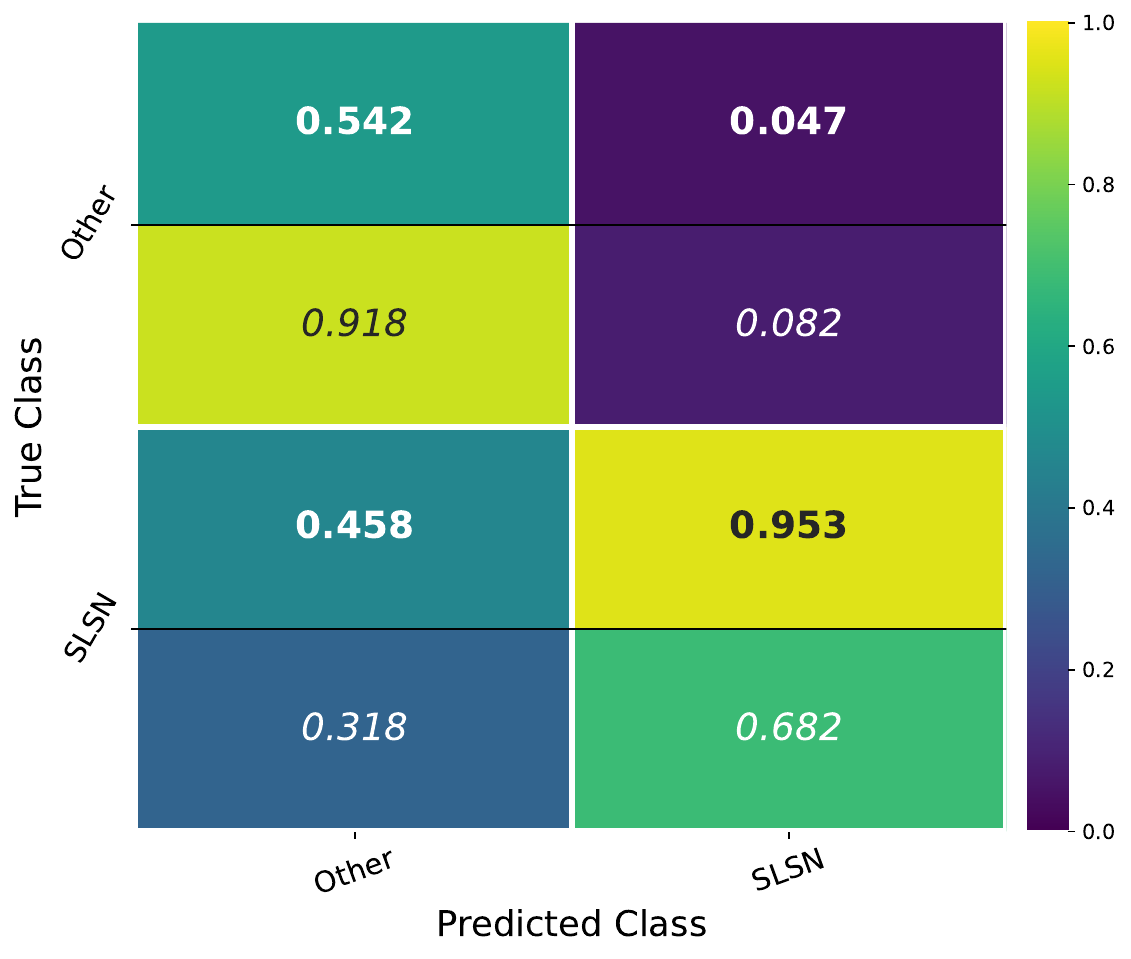}
    \caption{Performance metrics for the SLSN classifier applied to the sample of alerts classified as `long' by CATS.}
    \label{fig:cats_slsn}
\end{figure}
%%%%%%%%%%%%%%%
\par We take only the alerts that CATS classified as Long to assess the performance of the SLSN classifier. Results are shown in Fig. \ref{fig:cats_slsn}, where we can see improvement in both, precision (from 0.9 to 0.953) and recall (0.458 to 0.682), together with both AUCs (from 0.887/0.687 to 0.907/0.955 values for ROC/PR). Albeit small, this improvement could help produce more reliable follow-up catalogues with very few contaminants. The fraction of false negatives increased, simply due to the sample to be classified by the SLSN binary classifier being smaller.
\par The total completeness from this sample is approximately $35\%$, a small decrease when compared to the binary classifier alone. \rev{The fraction of correctly identified PISNe drops by approximately the same amount; a still significant amount of PISNe were identified by this hierarchical method, an important result considering the rarity of the class.}

This comes mainly from missing SLSNe, while the fraction of misclassified PISNe lowers by approximately $10\%$; an important result since PISNe are one of the least represented objects in the test set.  

\subsection{{\sc SuperNNova} binary into a broad classifier}
\label{subsec:snn_binary_combined}

Given that {\sc SNN} has one binary classifier for each broad class, we can combine them to create a multi-class classifier, potentially outperforming the broad {\sc SuperNNova} model. In order to obtain ROC and precision-recall curves, we join together all probabilities for all five binary broad classifiers and apply a softmax function to the probability vector, similar to what a standard DL multi-class classifier does. The predicted class is considered to be the one with highest probability. We note that there are a few light curves (approximately \rev{$0.09\%$}) that have all binary probabilities less than 0.5; that is, the models did not classify it as being in any of the broad classes. These alerts are mainly from the SN-like class, with SNII as the majority. We nevertheless assigned the broad class with the highest probability to them since it will not impact the results. 
\begin{figure}[ht!]
    \centering
    \includegraphics[width=\columnwidth]{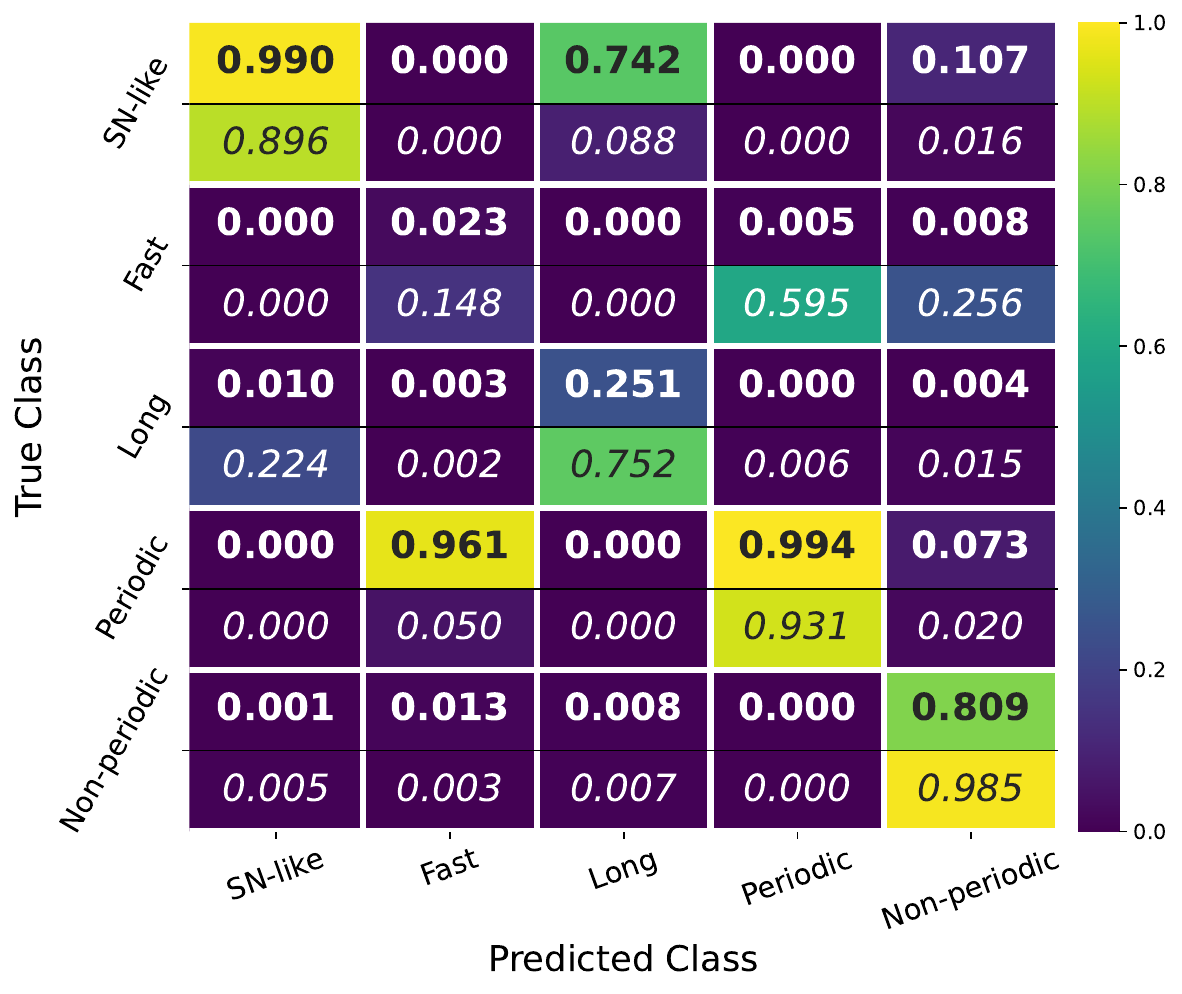}
    \caption{Performance metrics for the combined {\sc SuperNNova} binary classifiers.}
    \label{fig:combined_snn}
\end{figure}

Results are presented in Fig. \ref{fig:combined_snn} where we see \rev{improvements over the broad {\sc SuperNNova} classifier mainly for the SN-like and Long classes, while for the Periodic and Non-periodic the results are similar. The Fast class is the exception, where the combined binary results are worse than the broad {\sc SNN} and the binary {\sc SNN}; the reason is that for a lot of Fast alerts, the binary classifiers for Periodic and Non-periodic give larger probabilities than the Fast one. So although in a Fast versus others scenario they would be correctly classified, when combining all binary classifiers they end up being classified as either Periodic or Non-periodic, making the results worse.}  Since {\sc SuperNNova} requires balanced samples for training, the limitation on the number of alerts is applied on each class separately, instead of all classes being limited by the least represented. The model now is presented with more variety during training, and the probability of the correct class are improved.

\subsection{Combining classifiers for purity}
\label{subsec:making_pure}

Given the large volume of data LSST is expected to deliver on a daily basis, being able to build extremely pure samples out of tens of millions of light curves is extremely important, especially for follow-up purposes. Some of the models presented in Sect. \ref{sec:models} target the same class (or broad class); therefore, one way to improve the purity is to combine classifiers by only considering alerts that have been assigned the same class when presented to intrinsically different classifiers. 

\begin{figure}[hb!]
    \centering
    \includegraphics[width=\columnwidth]{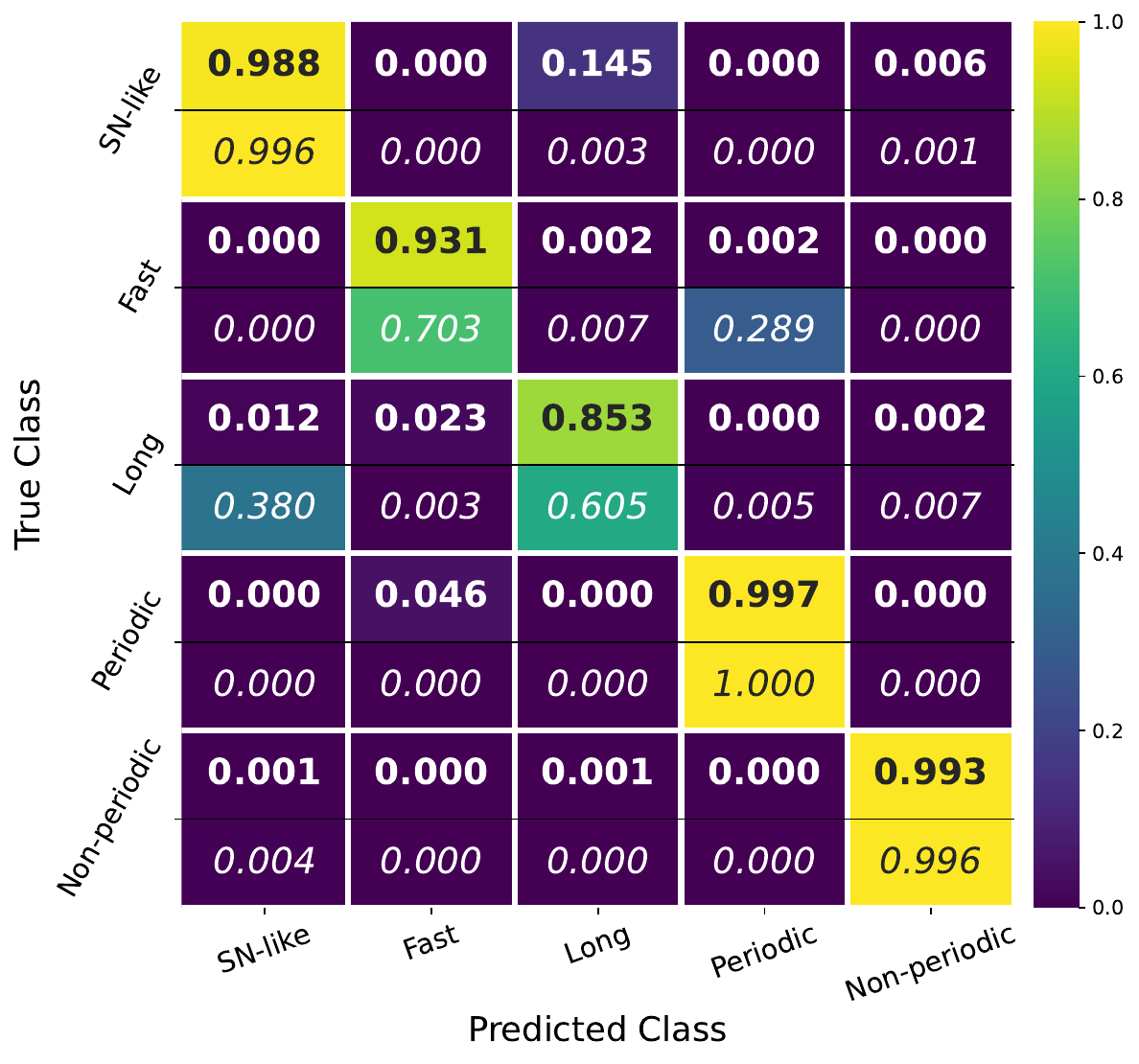}
    \caption{Confusion matrix for the alerts where CATS and each {\sc SuperNNova} binary classifier agree (see text for details).}
    \label{fig:cm_cats_snnbin}
\end{figure}

\rev{We analyse how our broad classification may be improved using CATS and the {\sc SuperNNova} binary classifiers: we only consider alerts for which the probability of the binary {\sc SNN} classifier corresponding to the class predicted by CATS is larger than $0.5$. This happens for approximately $96\%$ of the alerts in the test set.} We show the confusion matrix over this sample in Fig. \ref{fig:cm_cats_snnbin}, where it can be seen that the \rev{precision improved for all classes when compared to CATS and all {\sc SuperNNova} classifiers}. \revv{The recall is lower compared to {\sc SNN} for the long and fast classes (due to CATS).} 
%%%%%%%%%%
%%%%%%%%%%%%%%%%%%%%%%%%%
\section{Discussion}
\label{sec:discussions}
Our results show that all classifiers have mostly satisfactory performance for their classes of interest, with very few exceptions. Despite the difference in models and training methods in the classifiers, a few common trends emerged. 

All broad classifiers have trouble identifying the fast and long classes, for both the validation and test sets. CATS and {\sc SuperNNova} (both the broad and the combined binary classifiers) have different issues, with the former having higher purity and the latter higher completeness. Furthermore, looking at Figs. \ref{fig:cats_metrics}, \ref{fig:snn_broad_metrics}, and \ref{fig:combined_snn}, both CATS and {\sc SuperNNova} mix up the same classes: Long with SN-like, Fast with Periodic. This suggests that the similarity of the classes is intrinsic to the data set. 

The Periodic broad class is very well characterised in the data, with all multi-class classifiers achieving almost perfect purity and completeness. Indeed, the light curves for variable stars show a very distinctive periodicity, making them easy to identify among the other classes. 

CATS can be used as a first step in a hierarchical classification scheme, in which it sends to binary classifiers only the alerts belonging to their respective broad class. When applied to the SLSN classifier, it shows a small increase in purity, even though both classifiers were not meant to work together in the first place. The in-sample completeness \rev{shows significant increase when} compared to SLSN by itself.

The false positives of the EarlySNIa classifier are all within the SN-like broad class. This in turn means that using a broad classifier as a first step will not help improve its results.

{\sc SuperNNova} binary classifiers have better recall than the broad classifier for most classes, \rev{and an overall improvement particularly for the Long and SN-like classes.} Given the current configuration of {\sc SNN}, binary classifiers leverage better small training sets. Additionally, binary classifiers can be tuned to the target goal and improve performance with simple actions such as light curve length selection and hyper-parameter variation.

{\sc SuperNNova} binary classifiers can be combined into one multi-class model, with better (or similar, depending on the class) performance than its broad classifier. {\sc SNN} is lightweight, and thus it is feasible to do this without impacting LSST processing. A similar approach could be used by other broad classifiers, although care must be taken that the model is lightweight enough so that the extra computational cost does not affect the time necessary to process the alerts.

Samples with close to perfect purity can be built by requiring two or more classifiers to match their predictions. Combining CATS and all {\sc SuperNNova} binaries \rev{gives slightly better purity for all classes when compared to each individual model, with all classes except the Long reaching close to $100\%$ purity.}

\section{Conclusions}
\label{sec:conclusions}
% general ML

For a few years now, broker teams have been successfully working with the ZTF alert stream and communication protocols as a test bench for what is to be expected for LSST. This experience has been extremely successful and has allowed for the development of an entire broker ecosystem, along with a diverse and interdisciplinary community who supports it. Nevertheless, as valuable as this experience has been, it is also crucial to prepare the infrastructure for the important and challenging differences between the data delivered by the two experiments.

ELAsTiCC is a kind reminder that beyond hardware and data format, ML models and broker infrastructure will need to change significantly in order to fulfill expectations that will rise with the arrival of LSST. This includes the design of algorithms themselves, protocols for massive data transfers between geographically disconnected science teams, and experiment design for proper evaluation and optimisation of trained models to allow for the processing of millions of alerts per night. The analysis presented here describes the strategies developed by the {\sc Fink} team to address these issues. 

% ML algs
We introduced CATS, a deep learning classifier built especially to work with LSST data for broad classification, and it has shown great performance. Other classifiers, including {\sc SuperNNova} and tree-based models, were adapted from their current use on ZTF data. The adapted models performed well in their respective classification tasks, delivering pure and/or complete sub-samples. Moreover, we have also shown examples of how different algorithms can be used to build an ensemble classifier whose results outperform those from individual algorithms.

Nevertheless, it is important to keep in mind that {\sc Fink} operates in a framework where each classifier (science module) is developed independently by different science teams. Processing is centralised by the {\sc Fink} infrastructure, but model development is geographically and scientifically distributed. This means that each team has a different scientific goal in mind when developing their own classifier.

\par Given nine different classifiers (plus the combined ones) working at different tasks and showing different strong points, it is paramount to clarify the requirements for each science case so that the user can make an informed decision when 
applying the models presented here. CATS, for example, can produce pure samples for every class, but {\sc SuperNNova} \rev{(both multi-class and binary classifiers) have a better completeness for the fast and long classes and a similar purity for the SN-like and periodic classes.} In the future, it will be important to tailor models and/or combine different ones to achieve the top performance for a given science goal.
% training
\par We also call attention to details that should be kept in mind when interpreting the classification results stated here. \rev{First, we are bound by the diversity and complexity of the initial data set. The real data will certainly contain a number of intermediate objects whose properties lie in between classes, thus affecting these results. Moreover,} the choice of training sample is of utmost importance in ML problems regarding both their sample sizes and representativeness. Compared to its predecessor, PLAsTiCC, the objective of ELAsTiCC was the classification of alerts (i.e. partial views of complete light curves). Faced with the two different possibilities for a training set, we chose the streamed alerts since they most resemble the test set and results are more easily transferable to what is expected from LSST. 
%In addition, the split between training and a blind test set was a practical one. In the future, different experiment designs may be attempted to avoid the inherent biases that arise from  cutting the data in time.
% Data transfer
\par Moreover, it is crucial to recognise that training an ML model implies access to the data and to the necessary hardware to process it. During this challenge, a new service was designed by {\sc Fink} so that each team could access the curated data necessary for the training. This service was able to serve millions of alerts regularly to various teams, where the training of the models was largely performed on commodity hardware. Despite the undoubted usefulness of the data transfer service, irrespective of the volume of data to be transferred, we note that at the LSST scale, training models will require user teams to access dedicated hardware accelerators hosted on large data centres. This is an area where {\sc Fink} is actively planning on providing and thus enabling a service for the community to train models at scale.
% early classification
\par Considering the practical observational application of the classification results, early identification is paramount for the optimisation of follow-up resources and a major task of alert brokers. We have shown that most of our ML algorithms are capable of obtaining high-precision classification with less than 20 points. As more observations are added, the models generally give more accurate results. For the SN-like and fast classes, this performance increase is only valid with detection time spans related to their variability.

\rev{Finally, keeping in mind the intrinsic differences between simulated and real data, results presented here can be used to calibrate expectations regarding the output of \fink\ science modules in the first stages of LSST operations. Throughout the ten years of the survey, classifications will certainly evolve and present even better numerical results.}
% summary
\par In summary, we have shown that {\sc Fink} is able to process {\sc ELAsTiCC} alerts in a fast and efficient manner and provide informative ML classification scores for a variety of science cases. This proves the adequateness of the {\sc Fink} infrastructure and ML algorithms to process the big data volumes of Rubin LSST and, consequently, to contribute to the realisation of its scientific potential.

%%%%%%%%%%%%%%%%%%%%%%%%%%%%
\begin{acknowledgements}
We thank Robert Knop, Gautham Narayan, Rick Kessler and all others involved in the development of ELAsTiCC for enabling this work. We thank Konstantin Malanchev for adapting the \texttt{light\_curve} feature extraction package \citep[][\url{https://github.com/light-curve/light-curve-python}]{malanchev2021} for use within the EarlySNIa and SLSN classifier.  This work was developed within the {\sc Fink} community and made use of the {\sc Fink} community broker resources. {\sc Fink} is supported by LSST-France and CNRS/IN2P3. This is a result from the 2022 Fink Hackathon, 19 - 26 November 2022, Grimentz, Swizertland. This work received funding from CNRS MITI Evènements Rares - 2022, under project number 226696, Finding the first generation of stars with LSST (Fink). AM is supported by the ARC Discovery Early Career Researcher Award (DECRA) project number DE230100055. Part of this work was supported by the ARC Centres of Excellence OzGrav and OzGrav 2 project numbers CE170100004 and CE230100016. This work was supported by the `Programme National de Physique Stellaire' (PNPS) of CNRS/INSU cofunded by CEA and CNES. CRB acknowledges the financial support from CNPq (316072/2021-4) and from FAPERJ (grants 201.456/2022 and 210.330/2022) and the FINEP contract 01.22.0505.00 (ref. 1891/22). BMOF, CRB and AS acknowledge the LITCOMP/COTEC/CBPF multi-GPU development team for all the support in the artificial intelligence infrastructure and Sci-Mind’s High-Performance multi-GPU system. 
\end{acknowledgements}

\bibliographystyle{aa}
\bibliography{ref}

\begin{appendix}
\section{ROC and precision-recall curves}
\label{ml_metrics}

\begin{figure*}[ht!]
    \centering

    \subfloat[CATS]{\includegraphics[width=0.8\textwidth]{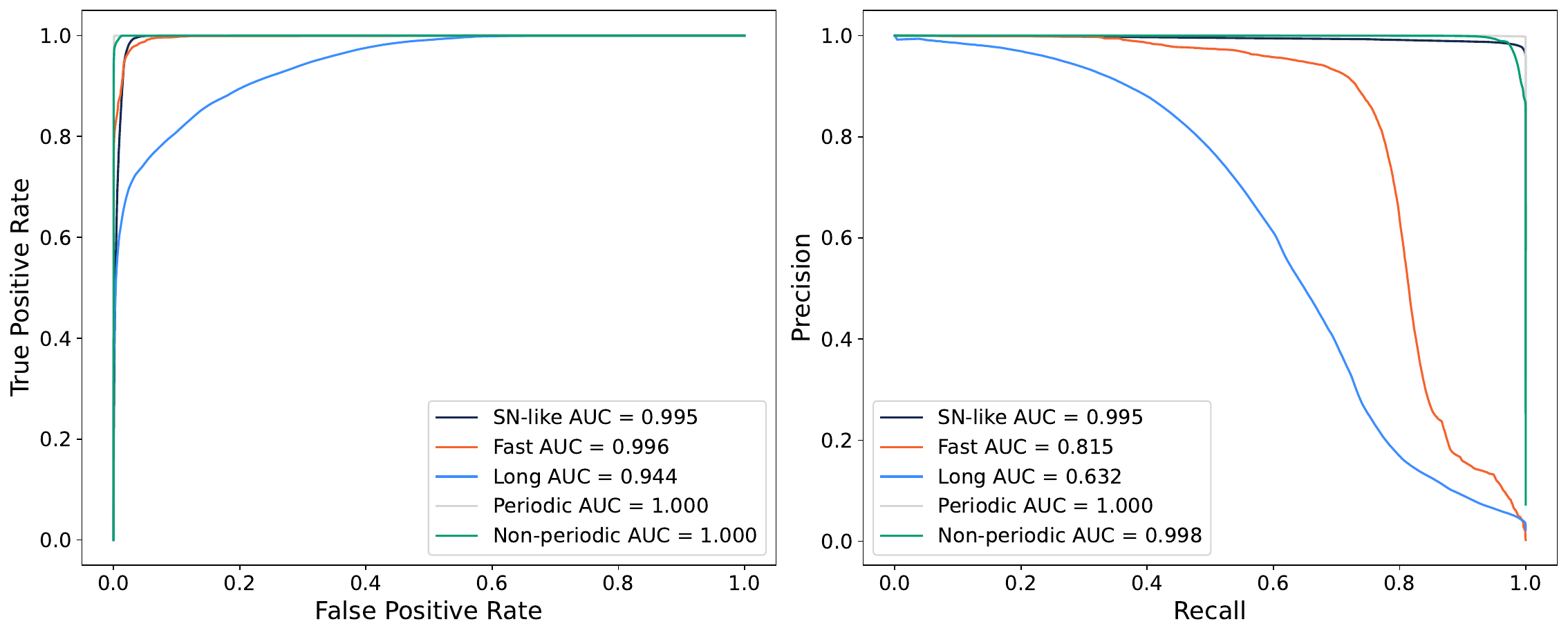}
    \label{fig:cats_curves}}

    \subfloat[{\sc SuperNNova} broad]{\includegraphics[width=0.8\textwidth]{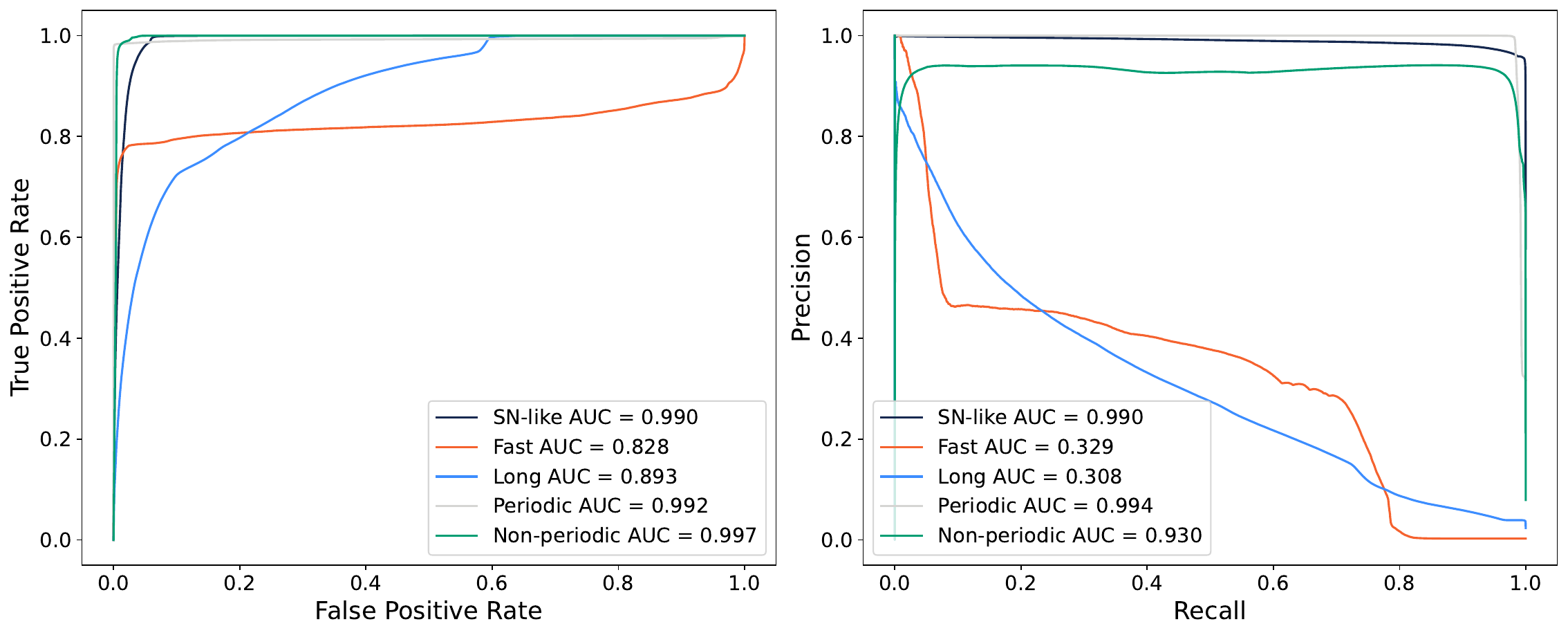}\label{fig:snn_broad_curves}}

    \subfloat[{\sc SuperNNova} binary classifiers]{\includegraphics[width=0.8\textwidth]{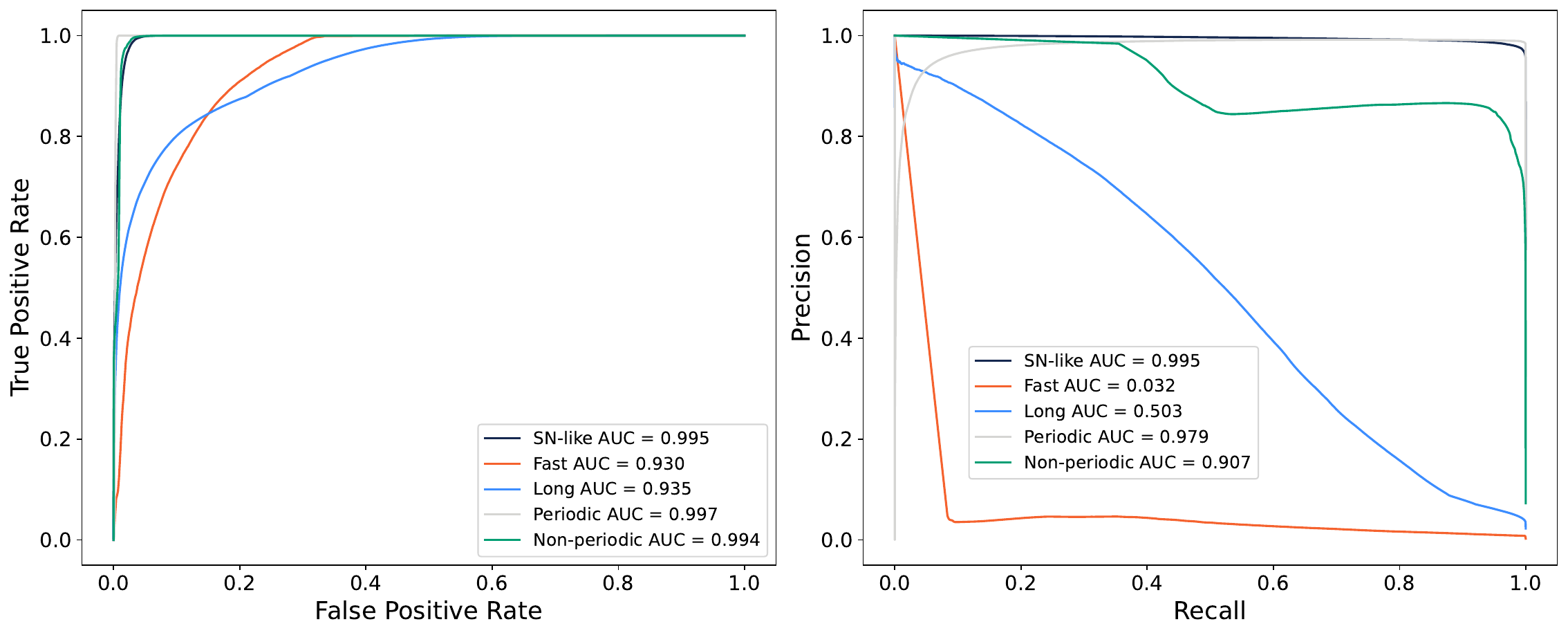}\label{fig:snn_bin_curves}}   
    \caption{Receiver operating characteristic and precision-recall curves for all classifiers. }
    \label{fig:ml_curves}
    
\end{figure*}

\begin{figure*}[ht!]{\ContinuedFloat}
    \centering

    \subfloat[SLSN]{\includegraphics[width=.8\textwidth]{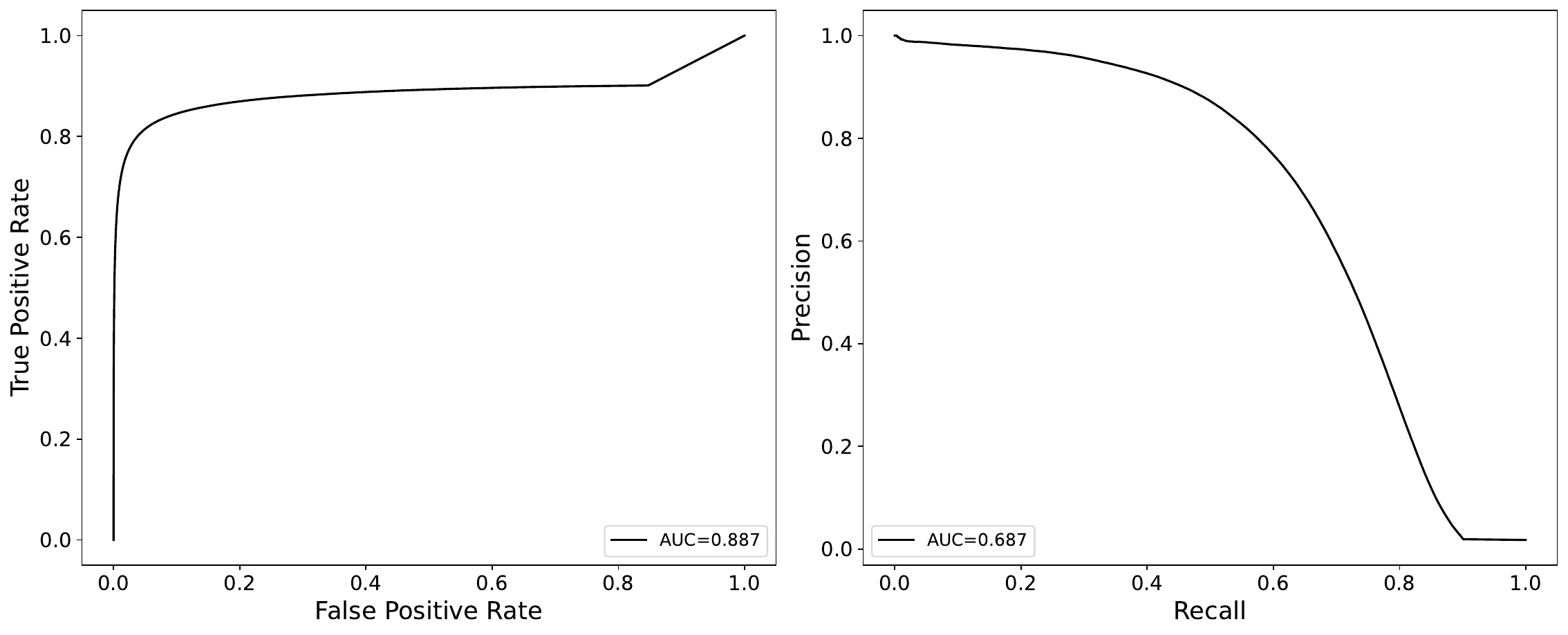}\label{fig:slsn_curves}}

    \subfloat[Early SNIa]{\includegraphics[width=0.8\textwidth]{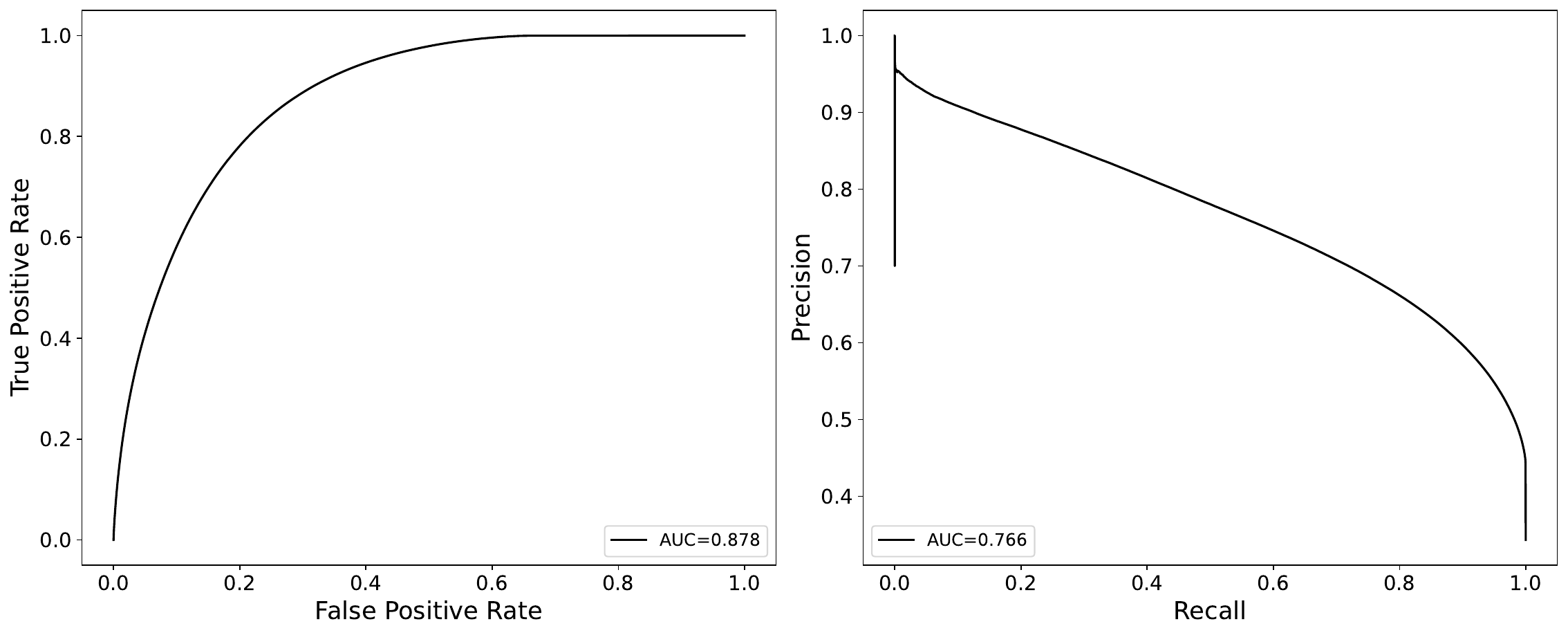}\label{fig:earlysnia_curves}}
    \caption{Continued.}
\end{figure*}
\FloatBarrier
\end{appendix}

\end{document}